\documentclass{aa}

\usepackage{graphicx}
\usepackage{txfonts}

\usepackage[usenames,dvipsnames]{color}
\definecolor{pinegreen}{RGB}{1, 121, 111}
\usepackage[draft, breaklinks=true,colorlinks,citecolor=pinegreen]{hyperref} 
\PassOptionsToPackage{unicode}{hyperref}
\PassOptionsToPackage{naturalnames}{hyperref}
\usepackage{natbib}
\usepackage{multirow}
\usepackage{xspace}
\usepackage{soul,xcolor}
\setstcolor{WildStrawberry}

\newcommand{\Rsun}{\ensuremath{\,\mathrm{R_\odot}}\xspace}
\newcommand{\Msun}{\ensuremath{\,\mathrm{M_\odot}}\xspace}
\newcommand{\Lsun}{\ensuremath{\,\mathrm{L_\odot}}\xspace}

\newcommand{\Myr}{\ensuremath{\,\mathrm{Myr}}\xspace}

\newcommand{\phiper}{$\varphi$~Persei\xspace}
\newcommand{\binaryc}{{\tt binary$\_$c}\xspace}
\newcommand{\MESA}{{\tt MESA}\xspace}

\newcommand{\AS}[1]{\color{WildStrawberry}\bf{#1}}
\newcommand{\SdM}[1]{{\color{Plum}\bf{#1}}}

\begin{document} 

   \title {Clues about the scarcity of stripped-envelope stars \\ from the evolutionary state of the sdO+Be binary system \phiper}
   
   \author{A. Schootemeijer 
            \inst{1,2}
            \and
            Y. G\"{o}tberg
            \inst{2}
            \and
            S. E. de Mink
            \inst{2}
            \and
            D. Gies
            \inst{3}
            \and
            E. Zapartas
            \inst{2}
            }
    \institute{ 
                Argelander-Instit\"{u}t f\"{u}r Astronomie, Universit\"{a}t Bonn, Auf dem H\"{u}gel 71, 53121 Bonn, Germany\\  \email{aschoot@astro.uni-bonn.de}
                     \and
                Anton Pannekoek Institute for Astronomy, University of Amsterdam, 1090 GE, Amsterdam, The Netherlands\\
                \email{Y.L.L.Gotberg@uva.nl, S.E.deMink@uva.nl, E.Zapartas@uva.nl}         
                \and
                Center for High Angular Resolution Astronomy and Department of Physics and Astronomy, Georgia State University, P.O. Box 5060, Atlanta, GA 30302-5060, USA \email{gies@chara.gsu.edu}\\
             }
             
    \authorrunning{Schootemeijer et al.}
    \titlerunning{Clues about the scarcity of stripped-envelope stars from the sdO+Be binary \phiper}

   \date{Received September -- ; accepted --}

   \abstract{ 
Stripped-envelope stars form in binary systems after losing mass through Roche-lobe overflow. They bear astrophysical significance as sources of UV and ionizing radiation in older stellar populations and, if sufficiently massive, as stripped supernova progenitors. Binary evolutionary models predict them to be common, but only a handful of subdwarfs with B-type companions are known. The question is whether a large population of such systems has evaded detection as a result of biases, or whether the model predictions are wrong. 
We reanalyze the well-studied post-interaction binary \phiper. Recently, new data improved the orbital solution of the system, which contains a $\sim$1.2\Msun stripped-envelope star and a rapidly rotating $\sim$9.6\Msun Be~star. We compare with an extensive grid of evolutionary models using a Bayesian approach and constrain the initial masses of the progenitor to $7.2\pm0.4\Msun$ for the stripped star and $3.8\pm0.4\Msun$ for the Be star. The system must have evolved through near-conservative mass transfer. These findings are consistent with earlier studies.
The age we obtain,  $57\pm9$\,Myr, is in excellent agreement with the age of the $\alpha$~Persei cluster. We note that neither star was initially massive enough to produce a core-collapse supernova, but mass exchange pushed the Be star above the mass threshold. %
We find that the subdwarf is overluminous for its mass by almost an order of magnitude, compared to the expectations for a helium core burning star. We can only reconcile this if the subdwarf resides in a late phase of helium shell burning, which lasts only 2-3\% of the total lifetime as a subdwarf.  Assuming continuous star formation implies that up to $\sim$50 less evolved, dimmer subdwarfs exist for each system similar to \phiper, but have evaded detection so far.  
Our findings can be interpreted as a strong indication that a substantial population of stripped-envelope stars indeed exists, but has so far evaded detection because of observational biases and lack of large-scale systematic searches.
}


   \maketitle
%
\section{Introduction \label{sec:introduction}}
Young stars of spectral type B and O are frequently found in binary and multiple systems \citep[for a review see][]{Duchene13}. Recently, it has become clear that the preference for very close binaries is stronger than generally assumed. Especially for stars at the high-mass end of the spectrum, interaction with a binary companion appears to be the dominant mode of evolution \citep{Kobulnicky06, Kiminki12, Sana12, Chini12, Almeida17}. In galaxies, these O-type and B-type stars and their more evolved counterparts play crucial roles in chemical evolution and feedback. They are the progenitors of white dwarfs, neutron stars and black holes that can give rise to many exotic phenomena, (explosive) transients, especially when they interact with a binary companion. Improving our understanding of the products of binary interaction is therefore relevant for multiple disciplines in astrophysics including the new field of gravitational wave astrophysics.

A large fraction of close binary systems evolve through a relatively simple first post-interaction phase. This phase starts after a first episode of stable Roche lobe overflow ceases. The original donor star has then lost nearly its entire hydrogen-rich envelope. The star becomes hot and compact and in most cases resides close to the helium main sequence \citep[e.g.,][]{Kippenhahn67, Pols91,Yoon10}. The less evolved companion still resides on the main sequence. Typically it now is the brightest star in the system, at least in the optical part of the spectrum. The stripped star emits most of its light in the far or extreme UV \citep[][]{Gotberg17}. This phase is usually relatively long-lived as both stars evolve on the slow nuclear burning timescale. We will refer to systems in this post-interaction phase as stripped-envelope star (SES) systems. The only way to avoid this phase is if the stars merge \citep[e.g.,][]{Podsiadlowski92} or experience certain proposed minority channels, for example those involving chemically homogeneous evolution \citep[][]{deMink09}, as proposed as a formation channel for binary black holes \citep[][]{Mandel16, Marchant16}. The  SES phase may be very short-lived for very wide systems where the donor fills its Roche lobe after completion of central helium burning (case C mass transfer).

Contrary to what might be expected for a common, long-lived phase, we only have sporadic identifications of SES systems. This poses an (apparent) paradox: despite the high binary fraction among young stars, these post-interaction binary systems containing an SES are rarely observed. This may be simply the result of observational biases, as argued for example by \cite{deMink14} and \cite{Gotberg17}. Post-interaction systems are very hard to detect due to a variety of selection effects and mostly pose as `apparently single stars'. While this may provide a solution for the paradox, the situation remains unsettling. Further insight into post-interaction systems and their detectability is warranted, especially because they can provide stringent tests for the uncertain mass transfer process. 

The few observed systems which contain a main sequence star + SES (e.g the Be + sdO  star systems, see Sect.~\ref{sec:implications}) clearly show how hard it is to detect them. The high temperatures of SESs hinder their detection due to the lack of absorption lines resulting from a high degree of ionization. Also, because of their high temperatures SESs shine most brightly in the UV: their companion typically outshines them in optical bands which are most accessible to observations \citep[e.g.,][]{Gotberg17}. Detecting their excess at short wavelengths requires observations from space, for example with the Hubble Space Telescope. These facilities are scarce and highly oversubscribed. Indirect evidence for their presence can be obtained if the main sequence companion shows radial velocity variations due to the orbital motion. Unfortunately these are often small because of an extreme mass ratio after mass transfer. Moreover, the main sequence star is expected to rotate rapidly \citep[e.g.,][]{deMink13} . Line broadening due to rotation as well as possible variations in the line profile due to the presence of an outflowing disk complicates the detection of small radial velocity variations.    

The Be + sdO binary system $\varphi$ Persei is one of the few exceptions for which the stripped remnant of a donor star has been detected. We elaborate on the observational history of this system in Sect. \ref{sec:persei}.  The sdO star is thought to be stripped of its envelope by the now rapidly rotating Be star \citep[e.g.,][]{Gies98} as has also been shown based on dedicated theoretical simulations for this system by \citet{Vanbeveren98, Vanbeveren98b} and \citet{Pols07}.  Recently, new data for this system have been presented by \cite{Mourard15} who unambiguously verified the presence of the SES by directly imaging the sdO component of \phiper using new high angular resolution observations.  Their astrometric data led to a slightly revised orbital solution, but consistent with those derived earlier by \cite{Gies98}.  The new observational data motivated us to reinvestigate the evolutionary status of the \phiper system and expand upon the theoretical studies presented earlier by \cite{Vanbeveren98, Vanbeveren98b} and \citet{Pols07} to find constraints on the physical processes.  In particular, we investigate the surprisingly high luminosity of the stripped star (its main sequence companion is eight times more massive, but only two times more luminous) and discuss the implications for its evolutionary status. We further place it in context of  analogue systems and use it to derive clues to better understand the apparent paradox of the scarcity of SESs.

The structure of this paper is as follows. In Sect. \ref{sec:persei}, we describe the observational history and characteristics of the \phiper system. To investigate the \phiper system we compute a large grid of binary evolutionary models with the fast binary evolution code \binaryc and selected models with the detailed evolutionary code \MESA, as is described in Section \ref{sec:method}. In Sect. \ref{sec:results} we present our main results. In Sect.~\ref{sec:modelvar} we discuss model variations, alternative implications and compare with previous studies. In Sect. \ref{sec:implications} we  give an overview of \phiper analogs that have been observed in different evolutionary phases. Sect. \ref{sec:conclusions} summarizes our conclusions.

\section{Observations of the \phiper system \label{sec:persei}}

At a distance of $183 \pm 3$ pc and an apparent magnitude of $m = 4.09$ \citep[][and references therein]{Mourard15}, \phiper can be observed with the naked eye. The main sequence star itself is classified as Be star of type B1.5 V:e-shell \citep{Slettebak82}: the `e' indicates spectral emission features and `shell' indicates that both rotationally broadened lines and narrow absorption lines are present, the second presumably originating from a circumstellar shell.  Already in the beginning of the 20$^{\textrm{th}}$ century radial velocity variations were detected and directly interpreted as a sign of binarity \citep{Campbell02, Ludendorff10, Cannon10}.

\begin{table}
\centering    
    \caption[]{ Parameters  for the subdwarf (sdO) and its companion (Be) in \phiper according to \cite{Mourard15} and references therein.  Of each component we list the present-day mass, $M$, effective temperature, $T_\mathrm{eff}$, luminosity, $L$ and surface gravity, $g$. Also, we list the  ratio of their radii, $R_\mathrm{sdO} \slash R_\mathrm{Be}$, the orbital period, $P_{\mathrm{orb}}$ and the eccentricity, $e$.
    
    }
    \label{tab:observed_paras}
\begin{tabular}{l c c}
\hline
\hline
            \noalign{\smallskip}
            & \textbf{sdO} & \textbf{Be}\\
            \hline
            \noalign{\smallskip}
            $M$ [M$_\odot$] & 1.2 $\pm$ 0.2 \hspace{0.82 mm} & 9.6 $\pm$ 0.3\\
            $T_\mathrm{eff}$ [kK] & 53 $\pm$ 3 \hspace{2.3 mm} & 29.3 $\pm$ 3 \hspace{3.3 mm} \\
            $\log_{10} ( L$ /  L$_\odot$)& $3.8 \pm 0.13$ & $4.16 \pm 0.10$ \hspace{-.2 mm} \\
            $\log_{10} (g / {\rm cm\,s^{-2}})$ & 4.2 $\pm$ 0.1 \hspace{0.9 mm} &  \\
            \hline
            \noalign{\smallskip}
            \noalign{\smallskip}
             $R_\mathrm{sdO} \slash R_\mathrm{Be}$        & \multicolumn{2}{c}{$0.20 \pm 0.01$} \\
            $P_{\mathrm{orb}}$ [d] & \multicolumn{2}{c}{$126.7 \pm 0.001$} \\ 
            $e$ & \multicolumn{2}{c}{$0$}\\
            \hline
\end{tabular}
\end{table}

The nature of the secondary has been matter of a long-standing debate.  \citet{Poeckert79, Poeckert81} reported the discovery of a weak He II $\lambda$4686 \AA ngstr\"{o}m emission line and found that it showed radial velocity variations moving in antiphase with the lines of the Be star primary. He argued that the emission line originated from hot gas surrounding the unseen companion. To explain the hard spectrum of the companion needed to ionize helium and produce the helium emission line, the author proposed that the companion could be a remnant core of a star stripped in a mass transfer event. \citet{Gies93}  discovered emission components of the He I $\lambda$6678 \AA ngstr\"{o}m line and found that they also moved in antiphase with the lines of the primary, supporting the findings by \citet{Poeckert81}. An overview of the available data at that time and a revised orbital solution was presented by \citet{Bozic95}

Motivated by the prediction that the still unseen companion would be bright in the UV, \citet{Thaller95} used the International Ultraviolet Explorer to search for direct features of the companion. 
They found that spectrum of the primary shows evidence of rotationally broadened photospheric lines, consistent with the expectation for a rapidly rotating Be star. 
They reported the discovery of many weak photospheric lines similar to those found in hot subdwarf O (sdO) stars as well as strong emission of C IV $\lambda$1550 \AA ngstr\"{o}m. Even though the spectrum was noisy (the hot companion only contributes about 12\% of the UV flux) they provided the first direct confirmation of the nature of the companion. \cite{Gies98} were able to confirm their findings by clearly showing the features of the faint hot subdwarf companion, using the Goddard High Resolution Spectrograph, which was one of the original instruments on board of the Hubble Space Telescope.

Recently, the full orbital solution of the \phiper system has been obtained through high angular resolution imaging observations in the near-IR and optical using long-baseline interferometry by \citet{Mourard15}, after combining their astrometric data with the earlier obtained radial velocity measurements. According to latest knowledge, the \phiper system consists of a very rapidly rotating ($v_\mathrm{rot} = 0.93 \pm 0.08$\,v$_\mathrm{crit}$) Be star with mass $M = 9.6\pm 0.3$\Msun and a hot SES of \textit{M} = $1.2\pm 0.2$\Msun with a 126 day orbital period, consistent with the parameters obtained earlier by \cite{Gies98}. Despite its much lower mass, the SES is almost half as luminous ($L_\textrm{SES}$ = 10$^{3.8 \pm 0.13}$\Lsun) as its Be star companion ($L_\textrm{Be}$ = 10$^{4.16 \pm 0.10}$\Lsun). The derived parameters of the \phiper system 
are displayed in Table \ref{tab:observed_paras}.

\section{Method \label{sec:method}}
 
To explore the parameter space with a large grid of evolutionary models we use the rapid synthetic binary evolution code  \binaryc, described in Sect. \ref{binaryc}. The advantage of this code is that it is fast enough to synthesize a large population of stars to use for our analysis. The downside is that certain physical processes are treated in less detail than in a full stellar evolutionary code. Therefore, we use the detailed stellar evolution code \MESA, described in Sect. \ref{mesa}, to test the evolutionary tracks of SESs that are predicted with \binaryc.

\subsection{Synthetic Binary Evolution with binary$\_$c \label{binaryc}}
This code was originally developed as a rapid single star evolution code \citep{Hurley00} based on analytical fits to the detailed stellar evolution models provided by \cite{Pols98}.  Later, it was made suitable to follow the evolution of binary systems, including physics such as tidal interaction and mass transfer as described in  \citet{Hurley02}. We use a version of the code, referred to as \binaryc developed and updated by \citet{Izzard04,Izzard06, Izzard09} and \citet{deMink13} to include more up-to-date physics assumptions.  
We adopt the same assumptions as listed in \citet[][and references therein]{deMink13} with the exception that we ignore rotationally enhanced mass loss via stellar winds. A full description can be found in the references above; here we give a summary of the physics assumptions that are most relevant in this work.
 
Overshooting is accounted for in the detailed evolutionary models that \binaryc relies on, using a parametrization for the overshooting parameter as described in \cite{Pols98}. In our mass range of interest it corresponds roughly to a classical overshooting parameter $\alpha_\mathrm{OV} \approx 0.28$, which refers to the extent of the mixed region above the convective core expressed in pressure scale heights. The assumed overshooting parameter is relevant since it determines the size of the convective core during the main sequence evolution and thus the mass of the resulting helium core. This in turn sets the mass of the stripped star after mass transfer.

Mass transfer and the criteria for stability of Roche lobe overflow are modeled as described in \citet{Hurley02}.  For post main sequence donors, mass transfer is assumed to lead to complete stripping of the hydrogen envelope. The further evolution is computed using fits against evolutionary tracks for naked helium star models \citep{Pols98}. 

We account for the rejuvenating effect of mass accretion using the algorithms described in \citet{Hurley02} but using an updated treatment for the core masses as described in \cite{deMink13}. This reduces the apparent age of the star while its mass increases, resulting in an increase of the temperature and luminosity of the accretor.  We also follow the angular momentum evolution of both stars and model the effect of spin up of the mass gainer as described in \cite{deMink13}.  We do not account for the possible additional effect of rotationally induced mixing processes or gravity darkening. We discuss possible concerns resulting from limitations in the treatment of rotation in section \ref{sec:results}. These affect the comparison with the rapidly rotating Be star. Our predictions for the stripped star, the main focus of this study, are not significantly affected by these limitations.

To allow for non-conservative mass transfer we limit the accretion rate $\dot{M}_\mathrm{acc}$ to ten times the mass of the accretor star $M_\mathrm{a}$ divided by its Kelvin-Helmholtz timescale $\tau_\mathrm{KH}$ (the time a star needs to radiate away its potential energy). If this limit is not exceeded, we assume that all mass lost from the donor star is accreted by its companion. Therefore we have:
\begin{equation}
\dot{M}_\mathrm{acc} = \mathrm{min}\left( - \dot{M}_\mathrm{donor}\, , \; \frac{10 M_\mathrm{a}}{\tau_\mathrm{KH}} \right)
\end{equation}
To investigate the impact of uncertainties in the mass transfer efficiency, $\beta = \dot{M}_\mathrm{acc} / \dot{M}_\mathrm{donor}$, we also consider cases where we vary  $\beta \in \{ 0.25, 0.5,0.75, 1.0\}$ following \citet{deMink07}. We thus assume implicitly assume that the accretion disk regulates the angular momentum that is accreted by the mass gainer; see discussion and appendix in \citet{deMink13}. We assume that all mass that is not accreted is lost from the system taking away the specific angular momentum of the orbit of the accretor star \citep[e.g.,][]{Petrovic05, vandenHeuvel17}

\subsubsection{The synthesized population \label{sec:synpop}}
With {\tt {\tt{binary$\_$c}}}, we evolve a grid of binaries until both stars in each system have left the main sequence. The following prior (birth) distribution functions are assumed:
\begin{itemize}
\item For the primary mass distribution,
\begin{equation}
f_{\textrm{M}_1}(M_1) \;d M_1 \propto M_1^\alpha \; d M_1. 
\label{eq:imf}
\end{equation}
The chosen mass range is $4.5\Msun \leq M_1 \leq 9.0\Msun$, which is large enough to find a solution for the \phiper system. An initial $M_1$ lower than $\sim$5.5\Msun is ruled out because the total mass of the system is $\sim$11\Msun and $M_2 < M_1$ by definition. The upper limit is based on a rough grid with a larger parameter space that was synthesized earlier, which showed no solutions at higher masses.  For this mass range we assume $\alpha = -2.35$, ergo a Salpeter initial mass function \citep{Salpeter55}.

\item The mass of the secondary star is drawn from the mass ratio ($q = M_2/M_1$) distribution,
\begin{equation}
f_q(q) \; dq \propto q^\kappa \; dq.
\end{equation}
This distribution is generally believed to be flat, see e.g., \cite{Kobulnicky06} and \cite{Kouwenhoven07}. 
 Therefore, we use $\kappa = 0$. We explore the mass ratio range $0.1 \leq q \leq 1.0$.

\item The chosen period distribution is flat in logspace, also known as \"{O}pik's law \citep{Opik24}: 
\begin{equation}
f_P(P) \; dP \propto \frac{1}{P} \; dP
\end{equation}
since $1/P \; dP = d\log P$. The chosen period range is 2.5\,days $\leq P \leq$ 40\,days. The limits are again based on the rough grid with a larger parameter space, which showed no solutions at lower and higher periods.
\end{itemize}
A flat prior probability distribution for the age of the system is used for the analysis.
The size of the grid is 24$\times$24$\times$24 grid (N($M_1$)$\times$N($q$)$\times$N($P$), 13824 systems in total).
We only consider circular systems: \cite{Mourard15} find that the current orbital solution converges a system with zero eccentricity. 
We assume a metallicity of $Z$ = 0.02 for the population. This metallicity matches the derived metallicity of the $\alpha$~Persei cluster \citep{Netopil13} with which \phiper is associated (see Sect. \ref{res:postpdf_mod}).

\subsubsection{Bayesian analysis \label{sec:bayes}}

We adopt a Bayesian approach that uses the prior probability of initial parameters to extract the posterior probability distribution functions using the synthesized population. Bayes' theorem can be expressed as:
\begin{equation}
P(M \mid D) = \frac{P(D \mid M)\; P(M)}{P(D)}
\end{equation}
where $M$ refers to a `model parameter' and $D$ refers to the `observed data'. $P(M\mid D)$ is the posterior probability distribution of an $M$ given the data from the observations ($D$). Here, we are interested in obtaining the posterior probability distributions of the following model parameters $M$: the birth masses of both stars, $M_\mathrm{SES, 0}$ and $M_\mathrm{Be, 0}$, the initial orbital period  $P_\mathrm{orb, 0}$ and the system age.
$P(M)$ is a model parameter's prior probability function, e.g., the initial mass function (eq. \ref{eq:imf}). $P(D)$ is not dependent on model parameters and serves as a normalization factor. 

$P(D \mid M)$ is the likelihood distribution of data points for a given a model parameter.
The observed parameters that we use to determine the likelihood of fit $P(D \mid M)$ of the model parameters are the masses and the temperatures of both stars in the binary, the orbital period and the ratio of the radii, which is a well-constrained value for the system. Therefore, with $M_\mathrm{SES}$, $M_\mathrm{Be}$,  $T_\mathrm{eff, \, SES}$, $T_\mathrm{eff, \, Be}$, $P$ and $R_\mathrm{SES} \slash R_\mathrm{Be}$
we use six fit parameters in total for the case of \phiper. Their values and errors are displayed in Table \ref{tab:observed_paras}.  The error on the observed orbital period is extremely small with $\sigma_\mathrm{P,obs} = 0.0071$\,days in comparison to the rather coarse spacing in the initial orbital period in our model grid. To prevent the orbital period from dominating the solution, we adopt instead an error of 15\,days. This corresponds to a relative error of about 10$\%$, which is similar to the relative errors on the other parameters \citep[cf.][]{deMink07}.

In practice, the method to obtain $P(M \mid D)$ is the following:
\begin{enumerate}
\item The likelihood of the fit to the observed parameters is calculated for each time step in each simulated binary system using the $\chi^2$ method. This corresponds to $P(D \mid M)$ and it is obtained using the following formula:
\begin{equation}
\label{eq:chi2}
\chi^2 = \sum\limits_{i} \left( \frac{x_\mathrm{i,obs} - x_\mathrm{i,mod}}{\sigma_\mathrm{i,obs}} \right)^2
\end{equation}
Where $x_\mathrm{i,obs}$ and $\sigma_\mathrm{i,obs}$ are the observed value and error for the fit parameter, respectively. The value of the fit parameter predicted by the model is represented by $x_\mathrm{i,mod}$.
\item The likelihood of the fit in each time step is then multiplied by the prior probability of the system, $P(M)$, which results from the probability distributions of the initial parameters $M_\mathrm{1, 0}$, $q_\mathrm{0}$ and $P_\mathrm{orb, 0}$, which are provided in Sect. \ref{sec:synpop}.
\item The value obtained at step 2 is multiplied by the time step and binned for the initial parameters $M_\mathrm{1, 0}$, $M_\mathrm{2, 0}$ and $P_\mathrm{0}$ and for the system age.
\end{enumerate}
The obtained result represents the posterior probability distribution of the system's initial parameters $P(M \mid D)$. In other words this is the probability distribution of the birth parameters.
We note that the error bars that we derive propagate from the observational constraints and do not take into account model uncertainties.

\subsection{Detailed Binary Evolution with MESA \label{mesa}}

The main focus of this study is the SES, or the sdO star in the system. The rapid synthetic evolutionary code described above treats these stars in a simplified way as pure helium stars. Using detailed evolutionary simulations \citep{Gotberg17} showed that these stripped stars typically still have a thin remaining layer of hydrogen, see also \citet{Yoon17}.  To investigate the consequences of this we also use the detailed binary evolutionary code \MESA \citep[version 7624,][]{Paxton11,Paxton13,Paxton15} to produce more realistic models for SESs.  This code solves for the one-dimensional stellar structure equations and composition changes simultaneously and follows the evolution of binary systems as they evolve through mass transfer.

We adopt the Ledoux criterion for convection, while using semi-convection with a value for the efficiency parameter of $\alpha_\mathrm{SC} = 1$ \citep{Langer83}. Mixing in the convective regions is described by mixing length theory \citep{Bohm58} and for the mixing length parameter we use $\alpha_\mathrm{MLT} = 2$. We adopt an overshooting parameter $\alpha_\mathrm{OV} = 0.28$ to be consistent with the models by \cite{Pols98} that are used in our rapid synthetic code \binaryc. We also take into account the effects of thermohaline mixing and rotation. The metallicity is again set to $Z = 0.02$. We use the nuclear network \texttt{approx21}, which contains 21 isotopes and is appropriate for modeling past central helium exhaustion. 

We account for mass loss through stellar winds, although the effect is very small in the mass range of interest. We use the algorithm by \cite{Vink01} for stars with $T_\mathrm{eff} > 10$\,kK and a surface hydrogen fraction of $X_\mathrm{H, s} > 0.4$. For hot stars that have lost part or most of their hydrogen-rich envelope, $T_\mathrm{eff} > 10$\,kK and $X_\mathrm{H, s} < 0.4$, we adopt the algorithm by \cite{Nugis00}. For cool stars, $T_\mathrm{eff} < 10$\,kK, the prescription from \cite{deJager88} is used. 
During stable Roche lobe overflow we use the implicit Ritter scheme \citep{Ritter88} to compute the mass loss rate from the donor star. Mass transfer is non-conservative as described in \citet{Gotberg17}, where mass loss from the system is assumed to take away the angular momentum of the accreting star.     We do not use MESA to further investigate the evolution of the accretor.




We simulate a binary system which has the same initial orbital period (16 days) and initial secondary mass (3.8 \Msun) as the best-fitting system we derive with \binaryc. The initial primary mass is chosen such that the SES has the desired mass after envelope stripping.

\section{Results \label{sec:results}}

Here, we discuss the results of our Bayesian analysis of the \phiper system against our grid of binary evolutionary tracks. To test our solution, we first describe the posterior distributions of fit parameters and compare them with the observed parameters (Sect.~\ref{res:postpdf_fit}) followed by a discussion of the properties of the progenitor system \phiper (Sect.~\ref{res:postpdf_mod}) and its inferred age  (Sect.~\ref{res:age}).  We then discuss the current evolutionary stage (Sect.~\ref{sec:evophase}) and its future evolution predicted by our best fit model (Sect.~\ref{sec:future}).

\subsection {Posterior distributions of the observed fit parameters \label{res:postpdf_fit}}
\begin{figure}
   \centering
   \includegraphics[width = \linewidth]{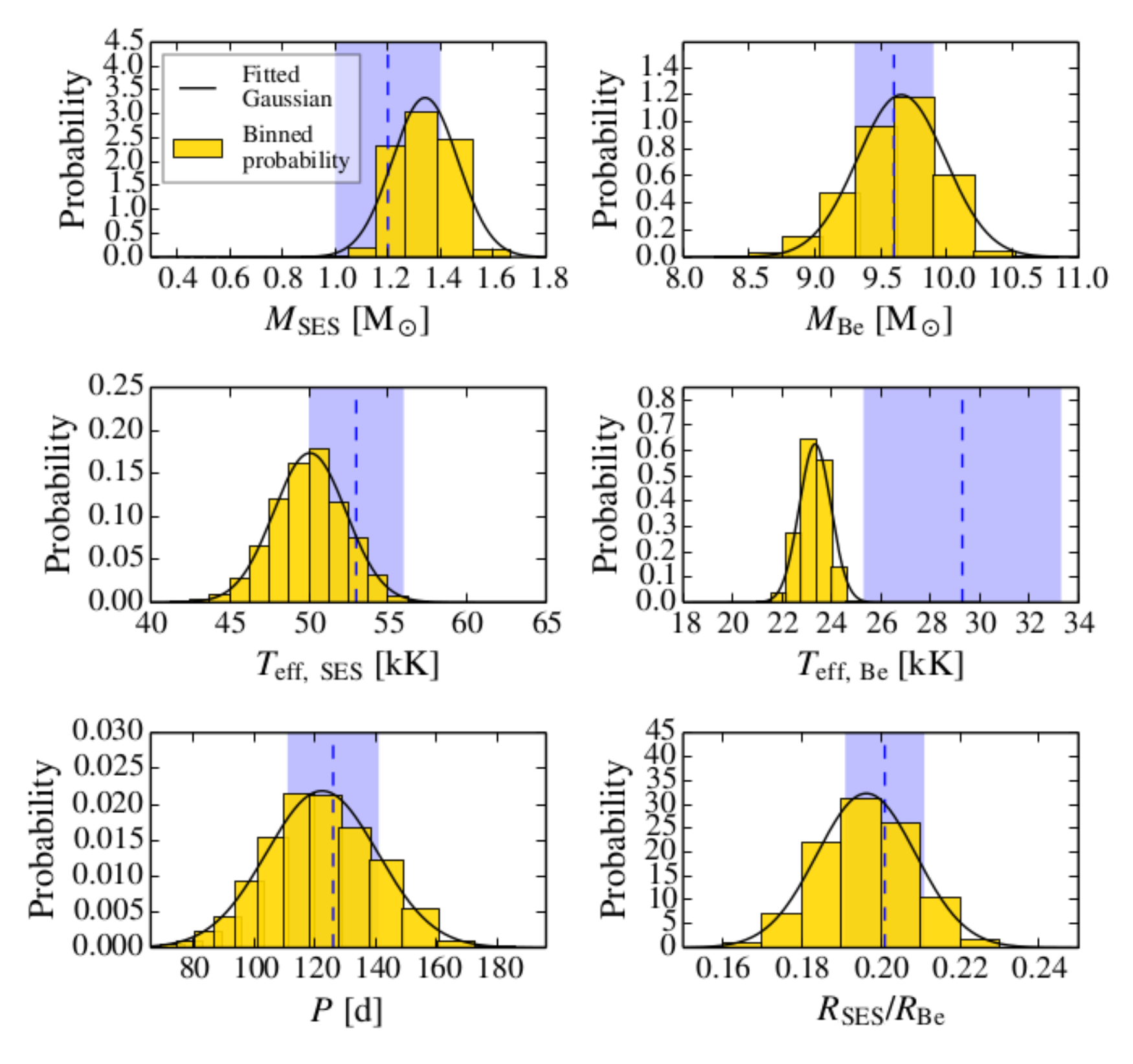}
   \caption{
   Posterior probability distributions of the fit parameters (yellow histograms) of the \phiper system together with a fitted Gaussian curve (black).
   For comparison, the observed parameters for the \phiper system (blue vertical dashed lines) and their 1\,$\sigma$ uncertainty intervals (blue shaded regions) are shown. 
   The top panels show the current masses of the sdO and Be star, $M_\mathrm{SES}$ and $M_\mathrm{Be}$. The central panels show their effective temperatures, $T_\mathrm{eff, \,SES}$ and $T_\mathrm{eff, \,Be}$.  The bottom panels show the present-day orbital period, $P_\mathrm{orb}$, and the ratio of their radii, $R_\mathrm{SES} \slash R_\mathrm{Be}$.
   }
              \label{fig:fit_paras}%
    \end{figure}

In this section we consider the posterior probability distribution of the fit parameters to test how well the models reproduce the observed system parameters. In an idealized case, both would be identical.
Fig. \ref{fig:fit_paras} shows a comparison of the observed parameters and their 1\,$\sigma$ uncertainty intervals with the posterior probability distribution of the fit parameters obtained from our analysis. The posterior probability distributions are well represented by a simple Gaussian, which indicates that they converge to a unique solution.


Generally we find good agreement between the observed parameters and the posterior probability distributions.  All peak within 1\,$\sigma$ of the observed value. The only exception is the effective temperature of the Be star,  $T_\mathrm{eff, \, Be}$. Our best-fitting models have temperatures that are 1.5\,$\sigma$ colder than the observed temperature of the Be star.  The difference may be of physical origin. It may indicate that the Be star has experienced extra mixing as a result of mass transfer. Stars that have a more homogeneous interior composition profile tend to be more compact and hotter.  Such mixing could for example result from thermohaline mixing if the Be star accreted helium rich material \citep{Stancliffe08} or rotational mixing \citep{Cantiello07} as a result of spin-up. In this case, these effects would more than compensate for gravity darkening \citep{vonZeipel24}. These are processes that we do not account for in these simulations. However, this does not significantly affect our findings for the evolutionary phase of the stripped object (see Sect. \ref{sec:evophase}), which is the main focus of this work.

\subsection{Constraints on the progenitor system of $\varphi$ Persei \label{res:postpdf_mod}}

\begin{figure}
   \includegraphics[width = \linewidth]{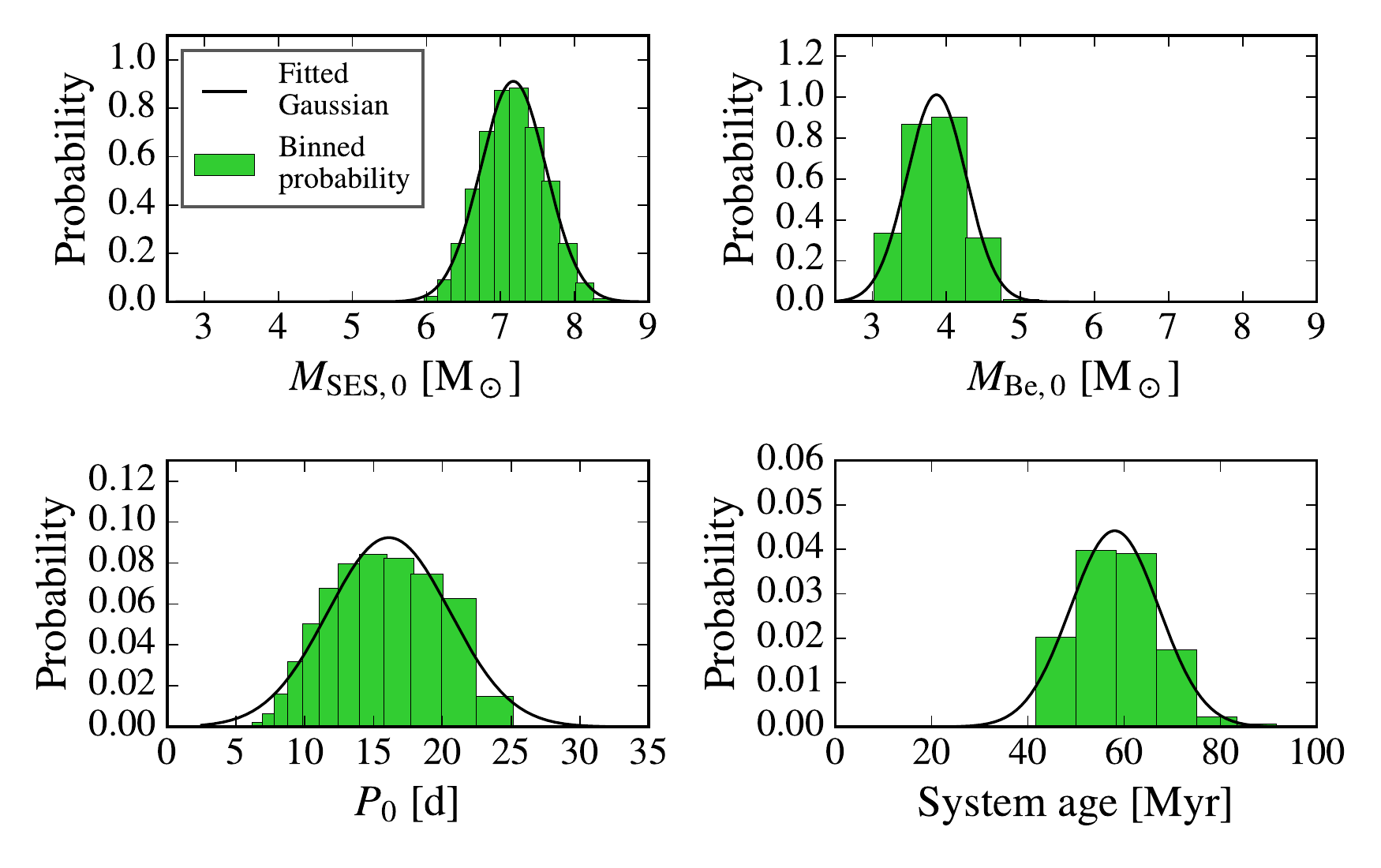}
   \caption{Posterior probability distributions of 
   the birth parameters (green histograms) of the \phiper system together with a fitted Gaussian curve (black). Top panels show the birth masses of the sdO star and the Be star,  $M_\mathrm{SES, \, 0}$ and $M_\mathrm{Be, \, 0}$. 
   Bottom panels show the initial orbital period, $P_\mathrm{orb, \, 0}$, and the system age.
   }
              \label{fig:initial_paras}%
    \end{figure}

In Fig. \ref{fig:initial_paras} we show the posterior probability distribution functions for the birth parameters derived using our default simulations. These are the initial masses masses of both stars, the initial orbital period and the current age of the system. The binned distributions are again well described by a simple Gaussian fit.

We find that the most probable initial masses of the system are $7.2 \pm 0.4\Msun$ for the primary star and $3.8 \pm 0.4\Msun$ for the secondary star. Interestingly, the present-day mass of the Be star is just above the threshold for a single star to end its life in a core-collapse supernova, $\sim 8\Msun$, while its birth mass was below this threshold.  This means that accretion  truly changed the final fate of the secondary  star in this system, which would have been destined to end its life as white dwarf if it were single.


For this initial orbital period, $16 \pm 4$\,days, the system was wide enough to remain detached during the main sequence evolution of the primary star. Mass transfer started after the initial primary started crossing the Hertzsprung gap, so-called case B mass transfer. 

We note that our findings are slightly different but consistent with those derived earlier by \cite{Vanbeveren98} and \cite{Pols07}.  They both inferred initial masses {\bf for the progenitors} of around 6\Msun for the SES and 5\Msun for the Be star, and an initial orbital period in the order of 10\,days. The differences are in part due to the fact that they tried to fit an sdO star mass of $1.14 \pm 0.04$\Msun \citep{Gies98}, which was the best estimate available at the time, while we use the more recently determined value of $1.2 \pm 0.2$\Msun instead.  
A further difference is that we simultaneously fit for multiple observed  parameters at once.  We do not fit the luminosity directly since this is a dependent parameter, but by fitting for the temperatures and ratio of the radii a fit solution that can explain the high luminosity of the sdO star is also favored -- this solution prefers a higher mass.

\subsection{The age of $\varphi$ Persei - a blue straggler of the $\alpha$ Persei cluster? \label{res:age}}

The age we derive for the \phiper binary system in the standard simulation is $57 \pm 9$\,Myr (Fig.~\ref{fig:initial_paras}).   \cite{Mourard15} argue that \phiper is likely a member of the $\alpha$ Persei cluster because its distance, kinematics and angular position are comparable to those of the cluster.  Our age estimate is in excellent agreement with recent estimates of the age of the $\alpha$ Persei cluster, which is in the vicinity of \phiper: these range from 52\Myr \citep{Makarov06} to 60\Myr \citep{Zuckerman12}. 

This age corresponds to a turn-off mass of about $\sim 7$\Msun star, which we consider the most likely initial mass of the original primary star in the \phiper system.  This by itself is supporting the picture that \phiper's Be star is a member of the cluster and that it gained a significant amount of mass from its companion later in its evolution and rejuvenated.  The Be star can be considered to be a rejuvenated blue straggler member of the cluster. 

 We note that the cluster age is debated: e.g., \cite{Stauffer99} found a cluster age of $90 \pm 10$\Myr using a method based on lithium depletion in brown dwarfs. The latter is not consistent with our findings. Also, it is on the high side of highest lifetime that the \phiper system could have given the minimum mass of the primary star (5.5\Msun, i.e., half the total system mass) which has a main sequence lifetime of $\sim$80\Myr.


   \begin{figure}
   \centering
   \includegraphics[width = \linewidth]{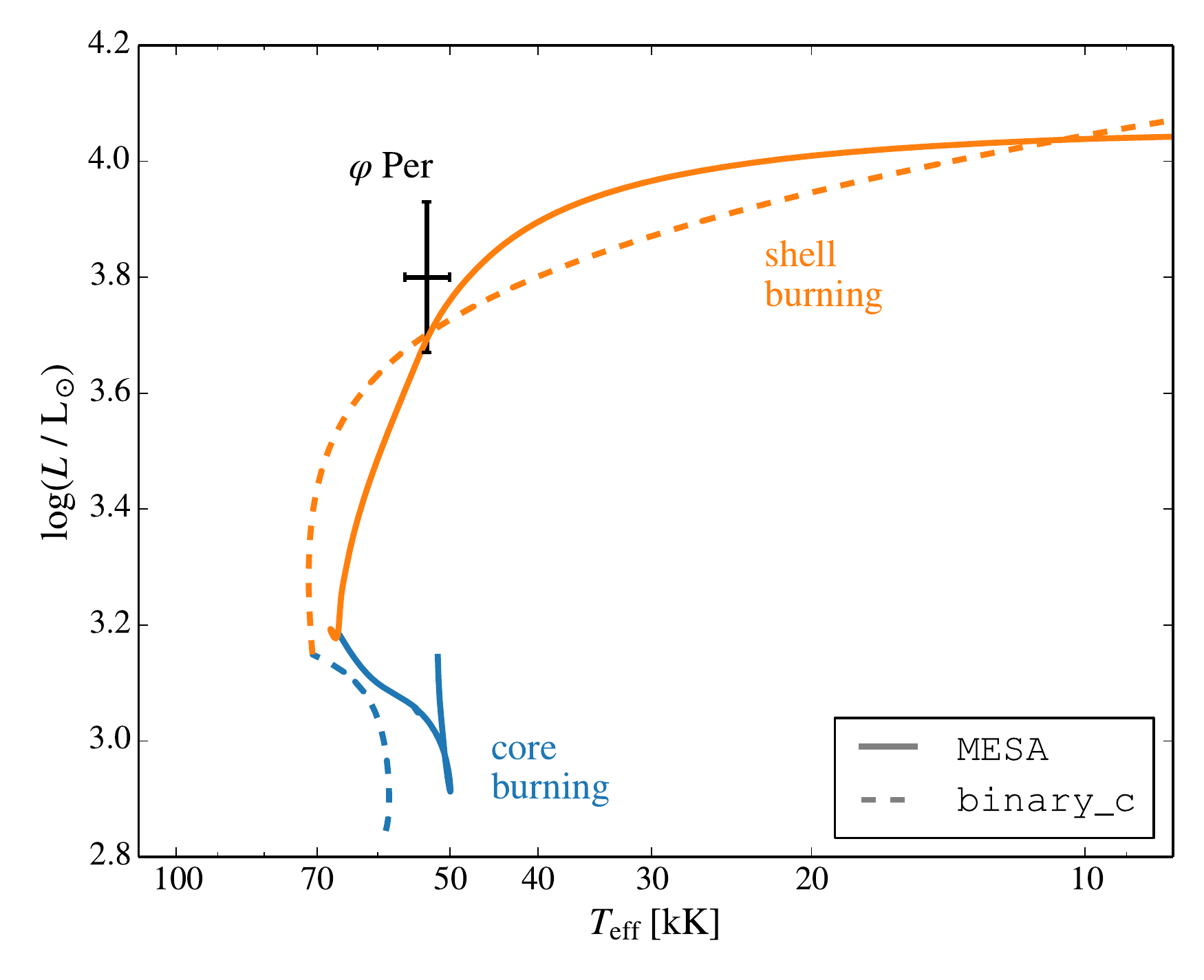}

   \caption{Hertzsprung-Russell diagram showing the evolutionary track of a stripped envelope star from the best-fitting system derived with {\tt{binary$\_$c}}. This object has a mass of 1.36\Msun. It is simulated with both {\tt{binary$\_$c}} (dashed line) and \MESA (solid line). Models in the helium core burning phase are displayed in blue; models in the helium shell burning phase are displayed in orange.  \label{fig:bestfit_track}}
   \end{figure}

\subsection{Evolutionary stage of \phiper -- caught in the short-lived helium shell burning phase \label{sec:evophase}}

We find that the best fit models for \phiper imply that the stripped star resides in the helium shell burning phase. This is illustrated in Fig. \ref{fig:bestfit_track}, where we compare the present-day parameters of the sdO star with the evolutionary track of our best fit model for the stripped star. We only show the stripped star phase after the first Roche lobe overflow has ceased. 
This includes the phase during which the star undergoes central helium burning. After exhaustion of helium in the core, burning continues in a shell. During this phase the star becomes more luminous and swells up.

In our \binaryc simulation the stripped star is treated as a pure helium star. For comparison we also show the corresponding evolutionary track of a stripped star resulting from a \MESA simulation. The \MESA model is slightly cooler during the helium core burning phase. This is due to a thin layer of hydrogen that is left at the surface, which sustains very weak burning of hydrogen in a shell around the core \citep{Gotberg17}. The \binaryc simulations ignore this thin hydrogen layer.  However, the differences between both tracks are small; they are comparable to the quoted observational error bars. We therefore conclude that our rapid \binaryc simulations are 
adequate 
for our analyses.

   \begin{figure}
  
   \centering
   \includegraphics[width = 
   \linewidth]{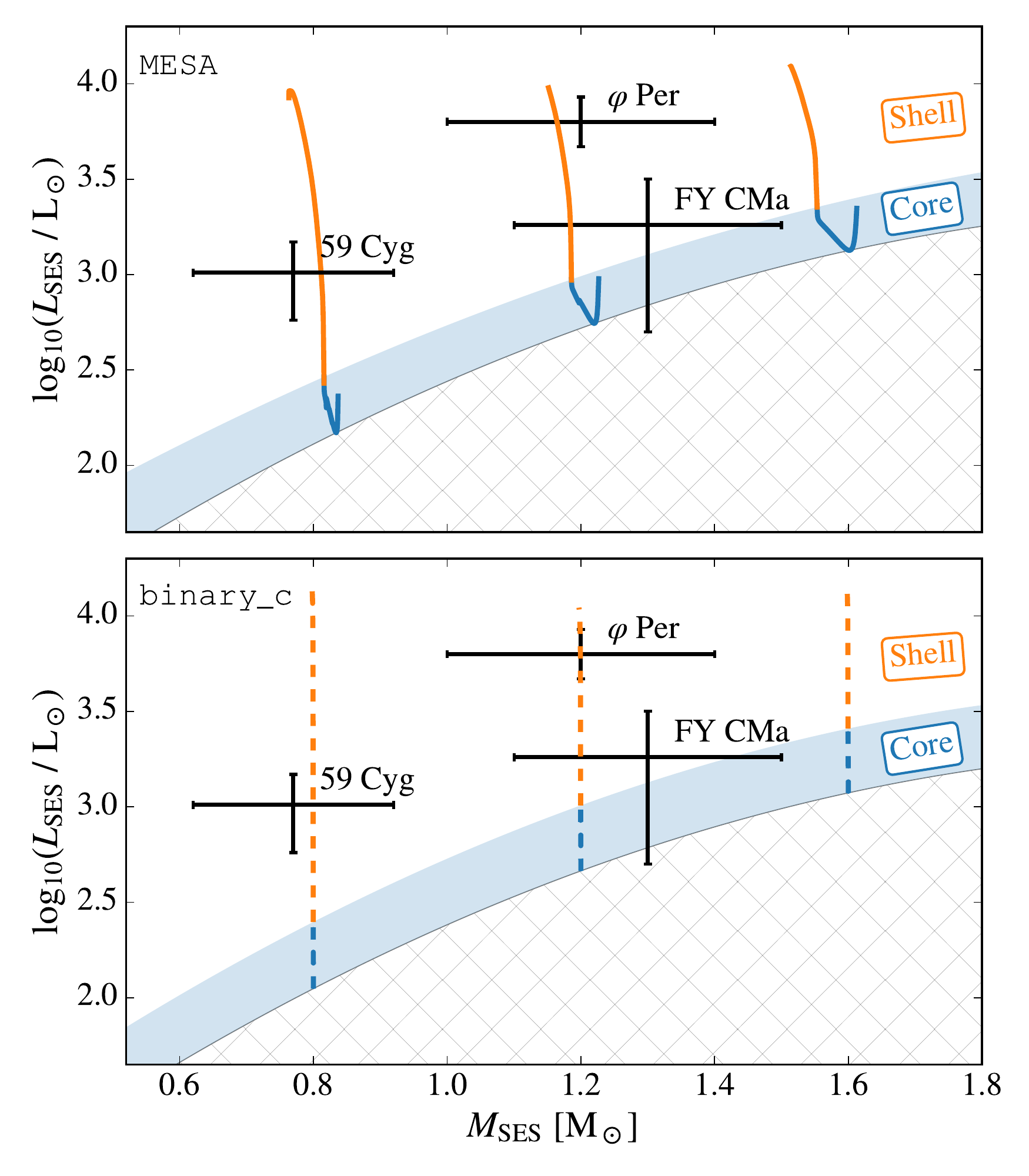}
   \caption{Evolution of the luminosity and mass of stripped envelope stars of 0.8\Msun, 1.2\Msun and 1.6\Msun. Helium core burning models are displayed in blue, while helium shell burning models are displayed in orange. The blue shaded area indicates the parameter space in which the models predict SESs to be helium core burning; in the area above, SESs are predicted to be helium shell burning. In the gray hatched region, the models do not predict helium burning SESs.
   \textbf{Top} (solid lines): \MESA tracks. \textbf{Bottom} (dashed lines): {\tt{binary$\_$c}} tracks.}
              \label{fig:Hestar_trax2}
    \end{figure}

Fig. \ref{fig:bestfit_track} thus indicates that the best-fitting models for \phiper's SES assume that it resides in the helium shell burning phase. The present-day luminosity is nearly 1\,dex brighter than the luminosity expected for models in the the helium core burning phase.  This is not only true for the best fitting model, but it holds for the full ensemble of likely solutions for \phiper. We find a posterior chance of $99.9\%$ that the system resides in the shell burning phase.

Fig.~\ref{fig:Hestar_trax2} further illustrates this by comparing observationally derived present-day mass and luminosity with the range of luminosities expected for helium core burning stars and helium shell burning stars, as a function of mass. For reference we show evolutionary tracks of 0.8, 1.2 and 1.6\Msun SESs simulated with both \MESA and  \binaryc. The effect of stellar wind mass loss is visible for the \MESA but apart from that we find that the differences 
are minimal. 
For both sets of models it is evident that \phiper is almost an order of magnitude too luminous to be helium core burning. Even if the mass of the object is 2\,$\sigma$ higher than the observationally derived value, the models strongly favor a helium shell burning solution.
 
To put the apparently high luminosity of \phiper's SES into perspective, we also show the parameters for the  two other known sdO $+$ Be binaries, FY\,CMa \citep{Peters08} and 59\,Cyg \citep{Peters13} in  Fig. \ref{fig:Hestar_trax2}. Their observed parameters are summarized in Table~\ref{tab:overview}.

Despite having the same mass within error bars ($1.3 \pm 0.2$\Msun) as \phiper ($1.2 \pm 0.2$\Msun), FY\,CMa's SES is about 0.5\,dex less luminous (Fig. \ref{fig:Hestar_trax2}). That is by itself surprising already. Because of this it fits to the helium core burning models as well as the helium shell burning models (Fig. \ref{fig:Hestar_trax2}).
It is remarkable that this object has been detected at all despite its luminosity being an order of magnitude lower than the luminosity of its Be star companion (Table \ref{tab:overview}). A possible explanation may lie in the extensive amount of UV observation time that has been dedicated to this system: between 1979 and 1995, 97 spectra were obtained with the International Ultraviolet Explorer satellite \citep[see][and references therein]{Peters08}. 
 
The SES in 59~Cyg has a lower mass ($0.77 \pm 0.15$\Msun) than the other two SESs. It seems, similar to \phiper's SES, to be too luminous for its mass to be in to the helium core burning phase by about 0.6\,dex, which is at over 2\,$\sigma$ (Fig. \ref{fig:Hestar_trax2}). 
This system seems to be a lower mass counterpart of \phiper, both being in the shell burning phase.


In summary we conclude that \phiper's SES is remarkably bright for its mass. It is about 0.5\,dex brighter than the stripped star in FY~CMa, which has the same mass within error bars. This is something we can only reconcile in our models if it resides in the short-lived helium shell burning phase.

\subsection{The future of the \phiper system - a possible progenitor of a late core collapse or exotic merger?\label{sec:future}}

According to our simulations, the primary star, which is now an sdO star, will end its life as a CO white dwarf. The secondary star, which is now a  $\sim$10\Msun Be star, is sufficiently above the threshold for single stars to end their life in a core-collapse supernova. Without interaction, neither of the components would have been massive enough to produce a core-collapse supernova. As such, \phiper is an example of a system that is expected to produce a  late core-collapse supernova \citep{zapartas17}. 

However, with a separation in the order of 200\Rsun, we expect the Be star (initially the secondary star) to fill its Roche lobe as it completes its main sequence evolution and swells up. This will initiate a phase of reverse Case B mass transfer from the secondary to the remnant of the primary, which is by then a hot, massive young white dwarf. Since the donor star is considerably more massive than its companion, the orbit is expected to shrink rapidly in response to mass transfer. The resulting phase of mass transfer may thus be unstable, resulting in a common envelope phase.  The outcome of this phase is uncertain. We expect that the most likely outcome is that the donor star engulfs the hot young white dwarf. The result is a merger with a peculiar interior composition.   In  \cite{zapartas17} we provide an extensive discussion and speculation of the possible outcomes of such a system.

\section{Discussion and model variations \label{sec:modelvar}}

In this section we discuss model variations, including variations in the mass transfer efficiency (Sect.~\ref{sec:beta}) and different cases of mass transfer (Sect.~\ref{sec:other_p0}), followed by alternative explanations for the excessive brightness of the sdO star in \phiper (Sect.~\ref{sec:altexpl}).

\subsection{A case for conservative mass transfer \label{sec:beta}}

Our default models assume a mass transfer efficiency that is governed by the thermal rate of the accreting star, as explained in Sec~\ref{binaryc}.  The mass transfer efficiency is however one of the primary uncertainties in binary evolution \citep[e.g.,][and references therein]{deMink07}. We therefore explore the impact of uncertainties in the mass transfer efficiency on the fit solutions.   We explore model variations where we adopt different constant values for the mass transfer efficiency $\beta \in \{ 0.25, 0.5, 0.75, 1\}$. This affects the best-fitting solutions and confidence intervals that we find for the initial masses, initial orbital periods and system age, as summarized in Table~\ref{tab:all_betas}.  The conservative models generally lead to solutions that have lower initial masses for both stars, larger initial orbital periods and larger system ages. 

Our default simulation and the simulations adopting a fixed value for $\beta = 1$ (conservative mass transfer) return identical solutions.  Models adopting conservative mass transfer yield better-fitting solutions --- we find  $\chi^2_\mathrm{min} = 3.77$ for $\beta = 1$ (fully conservative mass transfer), while the fit solution is  significantly worse for less conservative models with $\beta = 0.25$: $\chi^2_\mathrm{min} = 13.2$. 

The preference for conservative mass transfer to explain the system parameters of $\varphi$ Persei has been pointed out earlier by \cite{Pols07}. He used the mass of the sdO star to directly estimate the initial mass of its progenitor star assuming a moderate amount of core overshooting. The assumption that the progenitor of the sdO star was the initially more massive star of the system directly provides an upper limit to the initial mass of the secondary star.  By comparing the inferred initial masses with the present day masses, \cite{Pols07}  concluded that a mass transfer efficiency $\beta > 0.7$ is required, in agreement with our findings.

An independent argument in favor of conservative mass transfer comes from the inferred age. The \phiper system is a likely member of the $\alpha$ Persei cluster, which has an estimated age of 52-60\,Myr (see Sect. \ref{res:postpdf_mod}).  This is consistent with the ages we obtain when assuming relatively conservative mass transfer. The best fit ages obtained with our non-conservative models, $\beta \le 0.5$, differ by more than 1 sigma from the age of $\alpha$ Persei cluster. This further argues in favor of relatively conservative mass transfer, but it will be worthwhile to see whether Gaia data can give a more definitive answer concerning cluster membership.  


Finally, it is worth noting that our best fit models indicate a initial mass ratio $M_{2, \rm ini} / M_{1, \rm ini} \approx 0.5$. Evolutionary simulations of interacting binaries typically report that the accreting  star  in  such  systems  swells  significantly  during  the mass transfer phase \citep{Neo77} such that it may possibly even fill its Roche lobe too. If true, this would imply that the system briefly evolved through a near-contact or contact configuration. It has been suggested that such systems would experience significant mass loss from the system or even result in a stellar merger \citep[e.g.,][]{Pols94, Wellstein01, deMink07}. This is clearly in contradiction with the present-day appearance of the system. %
The current parameters of the \phiper provide a strong case that binary systems with similar initial mass ratios can evolve through a Roche lobe overflow phase  without any significant loss of mass from the system.

\begin{table}[]
\centering    
    \caption[]{ Derived birth parameters and age of the \phiper system for different assumptions of the mass transfer efficiency $\beta$, which we define as the fraction of the mass lost by the donor star that is accreted by the companion. Our default model assumes that the mass transfer efficiency is limited to ten times the thermal rate. Here we show the results for fixed values of $\beta$.  The conservative model, $\beta=1$ leads to identical solution as we have obtained in our default simulations. The best fitting solutions, yielding the lowest $\chi^2$ (Eq.~\ref{eq:chi2}) are found for models assuming conservative mass transfer.  
    \label{tab:all_betas}}
\begin{tabular}{l l l}
\hline
\hline

            \noalign{\smallskip}
             & Variable & Value\\
            \noalign{\smallskip}
            \hline
                      
            \noalign{\smallskip}
            \textbf{$\beta$ = 0.25} & $M_\textrm{1,ini}$ &  9.0 $\pm$ 0.4\Msun \\
             & $M_\textrm{2,ini}$ &  7.6 $\pm$ 0.3\Msun \\
             & $P_\textrm{orb,ini}$ & 6.3 $\pm$ 1.1\,d\\
             & System age & 36 $\pm$ 3\,Myr\\
             & Lowest $\chi^2$ obtained & 13.2\\
            \noalign{\smallskip}
            \hline

            \noalign{\smallskip}
            \textbf{$\beta$ = 0.5} & $M_\textrm{1,ini}$ &  8.1 $\pm$ 0.4\Msun \\
             & $M_\textrm{2,ini}$      &  6.3 $\pm$ 0.4\Msun \\
             & $P_\textrm{orb,ini}$ & 7.7 $\pm$ 1.6\,d\\
             & System age & 44 $\pm$ 5\,Myr\\
             & Lowest $\chi^2$ obtained & 7.14\\
            \noalign{\smallskip}
            \hline
            
            \noalign{\smallskip}
            \textbf{$\beta$ = 0.75} & $M_\textrm{1,ini}$ &  7.6 $\pm$ 0.5\Msun \\
             & $M_\textrm{2,ini}$      &  5.0 $\pm$ 0.4\Msun \\
             & $P_\textrm{orb,ini}$ & 10 $\pm$ 3\,d\\
             & System age & 50 $\pm$ 6\,Myr\\
             & Lowest $\chi^2$ obtained & 4.63\\
            \noalign{\smallskip}
            \hline
            
            \noalign{\smallskip}
            \textbf{$\beta$ = 1} & $M_\textrm{1,ini}$ &  7.2 $\pm$ 0.4\Msun \\
             & $M_\textrm{2,ini}$      &  3.8 $\pm$ 0.4\Msun \\
             & $P_\textrm{orb,ini}$ & 16 $\pm$ 4\,d\\
             & System age & 57 $\pm$ 9\,Myr\\
             & Lowest $\chi^2$ obtained & 3.77\\
            \noalign{\smallskip}
            \hline

\end{tabular}
\end{table}

\subsection{Different evolutionary scenarios \label{sec:other_p0}}

The only solutions we have found in our analysis as explanation for the present-day parameters of \phiper all involve stable Case B mass transfer, i.e., mass transfer where the Roche lobe filling star is crossing the Hertzsprung gap \citep{Kippenhahn67}.  

Even when we extend our grid of initial periods to include much wider systems, we do not find additional solutions. Such wide systems undergo late Case B or Case C mass transfer, where the donor star is a giant with a convective envelope at the onset of mass transfer. Mass transfer is unstable for these systems and leads to a common envelope phase.  The ejection of the envelope would lead to shrinking of the orbit, which could, in principle, with some fine tuning explain the present day orbital period.  However, we find that such solutions are incompatible with the present day masses. The initial mass of the donor star is effectively fixed by the present-day mass of the sdO star, since it sets the core mass of the original primary star.  This implies that the donor star initially cannot have been much more massive than about 7\Msun.  At the same time, solutions involving common envelope evolution imply that the  the secondary does not accrete any significant amount of mass.  This means that the present day mass of the secondary, which is now the Be star of $9.6\pm 0.3$\Msun sets a lower limit to the initial mass of the primary.  These two constraints are incompatible with each other. 

We also do not find solutions resulting from shorter period systems that undergo Case A  mass transfer (mass transfer from a main sequence donor).  Such solutions imply an initial period of $\sim$3.0\,days or less.  These are inconsistent with the large present-day orbital period of \phiper (126\,days), for any reasonable assumption of mass and angular momentum transfer.  This can be shown directly using the analytical initial-final period relation (equation 16.20 in \cite{Tauris06}, cf. \cite{Soberman97}), which require as input the initial and final mass ratio and the mass transfer efficiency $\beta$. For angular momentum loss from the system, we again assume that all escaping matter takes away the specific angular momentum of the secondary star (see Sect. \ref{sec:synpop}).  We explore the parameter space with primary mass $M_\textrm{1, \, 0} \, / \, \mathrm{M}_\odot \in \{ 7.2, 13.0 \}$, mass ratio $q_\textrm{0} \in \{ 0.0, 1.0 \}$ and mass transfer efficiency $\beta \in \{ 0.0, 1.0 \}$. For the current values of $M_1$ we use the stripped core mass which depends on $M_\textrm{1,\, 0}$. This stripped core mass is obtained by {\tt{binary$\_$c}}. The current $M_2$ follows from $M_\textrm{1, \, 0}$, $q_\mathrm{0}$ and $\beta$. We find that the lowest possible initial orbital period for which $P_\textrm{obs} = 126$\,days and $M_\textrm{2, \,obs} = 9.6$\Msun can still be produced is $P_0 \approx 4$\,days, which is larger than the maximum period for case A mass transfer at this mass.  From these results, we conclude that we were not able to find solutions for a case A mass transfer scenario.

We have not considered possibilities involving a triple companion, but we see no reason why this would provide an explanation of the overluminosity of the stripped star. 

\subsection{Alternative explanations for the excessive luminosity of the stripped star \label{sec:altexpl}}

An alternative hypothesis that one may put forward to explain the high luminosity of the stripped star in the \phiper's SES (with respect to its mass) is that the stripped star is currently accreting matter.
This is a possibility discussed by \cite{Mourard15}.   Even though the visible circumstellar disk of the Be star does not extend to the vicinity of the SES in the case of \phiper, this does not exclude the possibility of mass accretion onto the SES.  We use a simple estimate to assess the likeliness of this hypothesis.  

The stripped star appears to be almost an order of magnitude overluminous for helium core burning.  Disk accretion would thus be the dominant energy source of the SES. Adopting its currently observed mass and radius, we estimate the accretion rate that is necessary to reach its observed luminosity using the following expression,
\begin{equation}
    L_\mathrm{acc, \, SES} = \frac{G M_\mathrm{SES} \dot{M}_\mathrm{SES}}{R_\mathrm{SES}},
\end{equation}
and obtain an accretion rate of $\dot{M} > 10^{-4}$\Msun yr$^{-1}$.  This is several orders of magnitude higher than the typical mass loss rate of Be stars \citep[e.g.,][]{waters86}. We therefore deem this scenario implausible. 

Another possible hypothesis is that the stripped star exhibits a remaining layer of hydrogen and that hydrogen shell burning could be responsible for the high inferred SES luminosity.  A closer inspection of the detailed \MESA models shows that the fractional hydrogen luminosity is around 1$\%$ at the onset of helium shell burning, and it decreases further afterwards. The total hydrogen mass in the object at the onset of helium shell burning was 0.004\Msun. A much higher hydrogen mass than that would be needed for sustainable high luminosity as a result of hydrogen fusion. \cite{Gotberg17} show that even in the extreme case of very metal poor stripped stars, which retain a significant layer of hydrogen, the relative contribution of hydrogen burning luminosity is at best comparable to that produce by helium burning. We consider it unlikely that this can explain a order of magnitude discrepancy in the luminosity. A further reason to deem this scenario unlikely is that a larger hydrogen burning envelope would bloat the star and reduce its effective temperature. The \MESA model is already on the cold side of the observationally derived temperature.

%


\section{Implications of the short-lived evolutionary stage of $\varphi$ Persei for the population of undetected SESs \label{sec:implications}}

   \begin{figure}
   \centering
   \includegraphics[width = \linewidth]{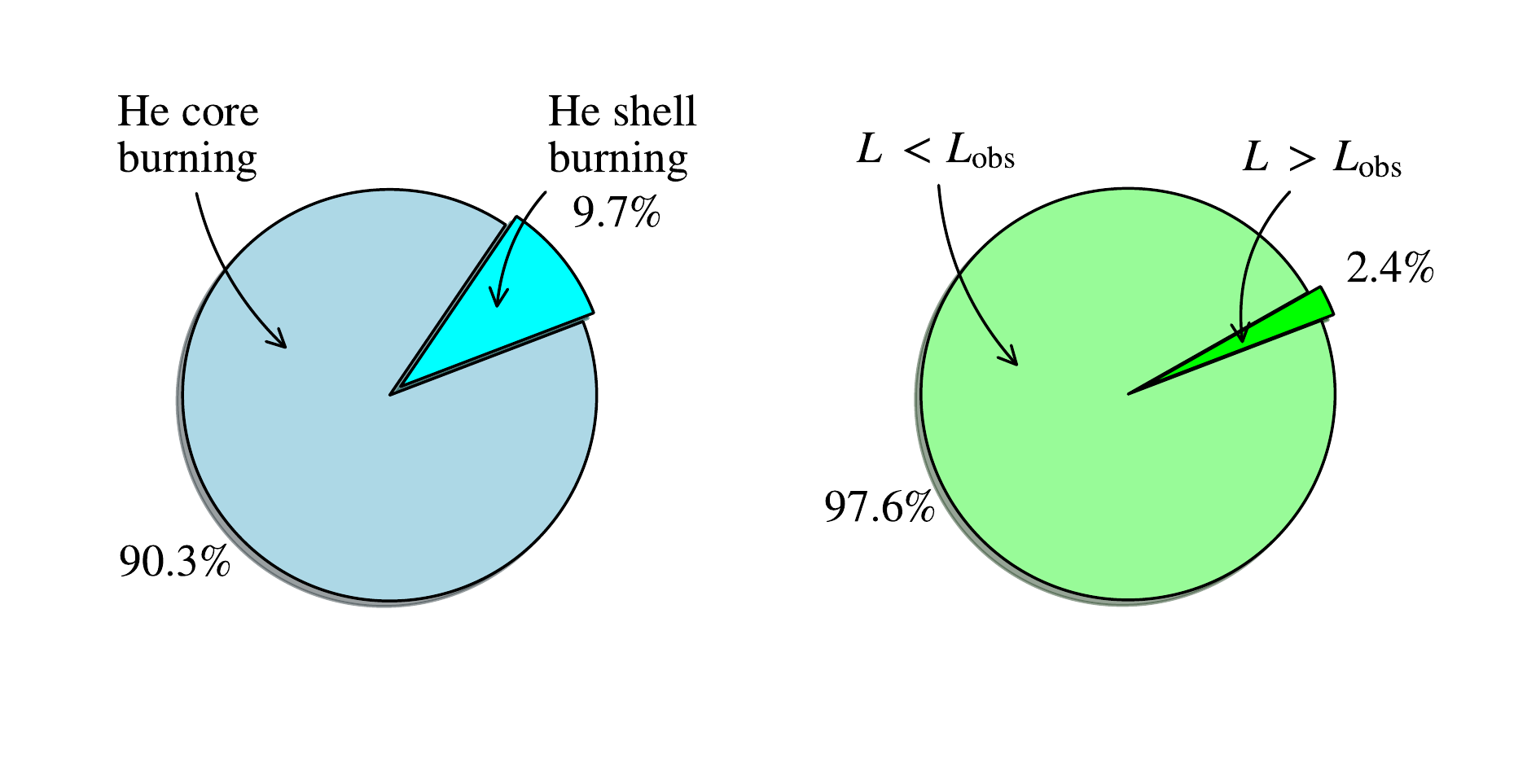}
   \caption{\textbf{Left}: fraction of time spent by SESs in systems similar to \phiper (see Sect. \ref{sec:evophase}) in the helium core burning phase (pale blue) and the helium shell burning phase (bright blue). \textbf{Right}: fraction of time spent by SESs in systems similar to \phiper below (pale green) and above (bright green) the luminosity limit $L_\mathrm{obs}$. Before adding up all the time steps in the grid to obtain the plots above, each time step is weighed by the likelihood of fit from their derived initial parameters.}
              \label{fig:pie_shell_lumi}%
    \end{figure}

Given the large fraction of young stars with nearby binary companions that are expected to interact during their lives one may naively expect SES system to be relatively common.  As a ball park estimate, we can assume that a fraction $f$ of all stars of intermediate mass are member of a binary system that will produce a SES. We further safely assume that the duration of the SES phase is about 10\% of the total lifetime, i.e., similar to the helium core burning lifetime. For any reasonable assumptions for $f$ and the assumption of continuous star formation, this leads to the expectation that in the order of few percent of all stars of intermediate mass should harbor a SES (see G\"otberg et al. 2018, submitted).  This ball park estimate appears to be in contrast with the very small number of systems that are currently known.  In the remainder of this section we discuss the clues that $\varphi$ Persei's evolutionary stage is giving us to understand this discrepancy.  

\subsection{How rare is  $\varphi$ Persei?}
In the previous sections we have argued that the SES in the $\varphi$ Persei system most likely resides in the short-lived evolutionary phase of helium shell burning.  The brightness of the SES in $\varphi$ Persei may have contributed in favor of its detection and characterization of an SES in such a short-lived phase.   For the full population of SESs, we expect the majority to reside in the long-lived helium core burning phase instead. Using our simulations we can estimate how many undetected systems we expect for each system like $\varphi$ Persei.

\begin{table*}[]
\centering    
    \caption[]{Observed parameters of HeGBs. The helium giant is labeled with `HeG'; the companion star is labeled with `c'. We elect to display the (to our knowledge) most recently published values. Values without errors are estimates. The letters in superscript refer to the studies where the values are adopted from (or references therein). These studies are: $^{a)}$\cite{Mourard15}, $^{b)}$\cite{Peters08}, $^{c)}$\cite{Peters13}, $^{d)}$\cite{Wang17}, $^{e)}$\cite{Peters16},
    $^{f)}$\cite{Drilling82}, $^{g)}$\cite{Kipper08}, $^{h)}$\cite{Parthasarathy90}, $^{i)}$\cite{Margoni88}, $^{j)}$\cite{Dudley90}, $^{k)}$\cite{Kipper12}, $^{l)}$\cite{Frame95}, $^{m)}$\cite{Schoenberner84}, $^{n)}$\cite{Dudley93}, $^{o)}$\cite{Jeffery01}, $^{p)}$\cite{Groh08}.
    Systems that are starred$^\star$ should have their masses divided by $\sin^3 i$. The (typically tiny) errors on the orbital period are omitted.}
    \label{tab:overview}
\footnotesize
\begin{tabular}{l l l l l l l l l l}
\hline
\hline
   \noalign{\smallskip}
    & $M_\mathrm{SES}$ & $T_\mathrm{eff\, SES}$ & $\log L_\mathrm{SES}$ & $\log g_\mathrm{SES}$ & $M_\mathrm{acc}$ & $T_\mathrm{eff \, acc}$ & $\log L_\mathrm{acc}$ & $P_\mathrm{orb}$ & $R_\mathrm{SES} / R_\mathrm{acc}$\\
     & [\Msun] & [kK] & [\Lsun] & & [\Msun] & [kK] & [\Lsun] & [d] & \\
     \hline
     \noalign{\smallskip}
     \textbf{Be+sdO:}\\
     \phiper & $1.2 \pm 0.2^{a)}$ & $53 \pm 3^{a)}$ & $3.80^{+0.13\, a)}_{-0.13}$ & $4.2 \pm 0.1^{a)}$ & $9.3 \pm 0.3^{a)}$ & $29.3 \pm 4.0^{a)}$ & $4.16^{+0.10, \, a)}_{-0.10}$ & $126.7^{a)}$ & $0.20 \pm 0.01^{a)}$\\
     \noalign{\smallskip}
     FY CMa & $1.3 \pm 0.2^{b)}$ & $45 \pm 5^{b)}$ & $3.26^{+0.26 \, b)}_{-0.56}$ & $4.3 \pm 0.6^{b)}$ & $11.5 \pm 1.5^{b)}$ & $27.5 \pm 3.0^{b)}$ & $4.26^{+0.26 \, b)}_{-0.60}$ \hspace{-50em} & $37.3^{b)}$ & $0.12 \pm 0.04^{b)}$\\
     \noalign{\smallskip}
     59 Cyg & $0.77 \pm 0.15^{c)}$ & $52 \pm 5^{c)}$ & $3.01^{+0.16 \, c)}_{-0.24}$ & $5.0 \pm 1.0^{c)}$ & $7.8 \pm 1.5^{c)}$ & $21.8 \pm 0.7^{c)}$ & $3.91^{+0.10 \, c)}_{-0.14}$ & $28.2^{c)}$ & $0.062 \pm 0.003^{c)}$\\
     60 Cyg & $1.7^{d)}$ & $42 \pm 4^{d)}$ & & & $11.8^{d)}$ & $27^{d)}$ & & $146.6^{d)}$ & $0.093 \pm 0.012^{d)}$\\
     HR 2142 & $0.7^{e)}$ & $ \geq 43 \pm 5^{e)}$ & $> 1.7^{e)}$ & & $9^{e)}$ & & & $80.9^{e)}$ & $> 0.026^{e)}$\\
     \hline
     \noalign{\smallskip}
          
     \textbf{HeGBs:}\\
     KS Per & $1^{f)}$ & $9.5 \pm 0.4^{g)}$ & $3.3^{g)}$ & $2.0 \pm 0.5^{g)}$ & $5-6^{f)}$ & $30^{h)}$ & & $362.2^{i)}$\\
     $\upsilon$ SGR$^*$ & $2.52 \pm 0.10^{j)}$ & $12.3 \pm 0.2^{k)}$ & $ 4.6^{j)}, 3.7^{k)}$ & $2.5 \pm 0.5^{k)}$ & $4.05 \pm 0.05^{j)}$& & & $138^{l)}$\\
     LSS 4300 & $1^{m)}$ & $12 \pm 1^{n)}$ & $ 4^{m)}$ & $1.4^{m)}$ &  &  &  & $51.1^{l)}$\\
     LSS 1922 &  & $12 \pm 0.5^{n)}$ &  &  &  &  &  & \\
     BI Lyn & $0.5^{o)}$ & $28.6 \pm 1.0^{o)}$ & $ 3.32^{o)}$ & $3.6 \pm 0.1^{o)}$ &  & $5.9 \pm 1.0^{o)}$ &  & \\
     \hline
     \noalign{\smallskip}
          
     \textbf{qWR+B:}\\
     HD 45166 & $4.2 \pm 0.7^{p)}$ & $50 \pm 2^{p)}$ & $3.75^{+0.08 \, p)}_{-0.08}$ &  & $4.8 \pm 0.5^{p)}$ & $13.5^{p)}$ & & $1.6^{p)}$\\

\end{tabular}
\end{table*}

We quantify this using our model grid by providing an average for the ensemble of systems that have  birth parameters similar to those of \phiper.  To obtain this we compute the relative duration of helium shell burning for all SESs in our grid and weigh them with their likelihood of the fit to the derived initial parameters of \phiper.  
We find that the helium shell burning phase lasts on average around 10$\%$ of the total helium burning lifetime (Fig. \ref{fig:pie_shell_lumi}, left).
During this phase the SES increases in brightness.  In a similar fashion we can compute the average time that SESs in systems with birth parameters similar to \phiper are as bright as or brighter than \phiper's SES. We find that the time a SES born in systems with similar birth parameters as $\varphi$ Persei spends only $2.4^{+0.9}_{-0.7}\%$ of their SES lifetime having a luminosity $L > L_\mathrm{obs}$ (Fig. \ref{fig:pie_shell_lumi}, right). The error bars here reflect the uncertainty in the observed luminosity.  We obtain a very similar result if we analyze the \MESA evolutionary track instead, which has a luminosity of $L > L_\mathrm{obs}$ for $2.7^{+2.1}_{-1.3}\%$ of its lifetime.

The stripped star in \phiper thus belongs to the top $\sim $2-3\% brightest in its class. This implies that for each system similar to \phiper, with a stripped star that is as bright or brighter than the stripped star in \phiper, we expect about 30-50 systems hosting a dimmer stripped star, most of which are residing in the helium core burning phase. Their presence will be very hard to detect through their luminosity alone. This, by itself, is providing insight in the observational difficulties to  detect stripped stars. 

 \subsection{The $\varphi$ Persei system in context of the known SESs}

To try and place  $\varphi$ Persei in context we have conducted a literature study and have attempted to compile a complete overview of all currently known SESs with 
main sequence companions.
We have only been able to find eleven systems matching these criteria\footnote{We chose not to include classic Wolf-Rayet stars here nor sdO and sdB stars that have dim, low mass ($\sim$1\Msun or less) white dwarf or late-type main sequence star companions. We refer to \citet[][]{Kupfer15} and \citet{Heber16} for an overview of these.}. This includes $\varphi$ Persei as well as the systems FY\,CMa and 59 Cyg to which we already compared in Fig.~\ref{fig:Hestar_trax2}. Unfortunately, in most cases several of the parameters are still unknown or only contradictory estimates are available.  We provide of the data that we have been able to compile in Table~\ref{tab:overview}.  

We found a total of five systems that appear to be systems hosting a subdwarfs plus a Be star (Be+SdO).  Aside from the three systems mentioned above, the recently detected SESs in  60 Cyg \citep{Peters16} and HD2142 \citep{Wang17} belong to this group.  They all host subdwarfs with masses around 1\Msun, temperatures of 40--50\,kK, high effective gravities of $\log g = 4.2$--5.  Their companions have masses ranging from 8--12\Msun and orbital periods ranging from 1--5 months.  All these appear to be post-interaction systems resulting from stable and rather conservative mass transfer.  

We found an additional set of five systems that have been marked as helium giants (see \cite{Dudley93} and references therein, and \cite{Jeffery01}). These are thought to be helium stars that are in a very late phase of helium shell burning. 
The expectation is that these binaries are currently going through or are about to go through their second mass transfer phase \citep[e.g.,][]{Schoenberner83} as is predicted by \cite{deGreve77} and \cite{Delgado81}.
In this case, the donor star fills its Roche lobe again during a late helium shell burning phase in which it cools and expands. The SESs in these systems have lower temperatures, 9-30\,kK, and lower effective gravities of  $\log g = 1.5$--3.5.  Note that several the parameters for these system result from studies conducted in the 1980s. Often the nature of the companion is not known or and/or its presence is only inferred from radial velocity variations.  Observing these systems again would be worthwhile, to try and obtain more reliable measurements. One of these stands out, $\upsilon$ Sgr, which is the most massive one in this category.  The SES is a possible progenitor of a core-collapse supernova. 

We have added one additional system to this compilation that does not belong to either of these groups. This is the quasi-WR star in HD 45166, which has a late B-type companion in a tight orbit of 1.6 days. Also worth mentioning is that  \cite{Smith17} and \citet{Gotberg17} recently argued that the new class of WN3/O3 stars discovered by \citep{Massey14, Massey15, Massey17} may also be SES stars.  At least one of these objects is a short period eclipsing binary system \citep{Graczyk11}. Whether all of these stars have companions is still topic of debate. 
 
It is striking that the majority of the SESs in these systems appear to be in their helium shell burning phase.
This includes all five HeGB systems, $\varphi$ Persei and 59 Cyg
.  That is a total of at least seven out of a sample of eleven. For each of these seven we expect about 10-50 additional systems to still harbor undetected dimmer SESs.  This seems to suggest we are missing hundred(s) of SESs in the sample from which these binary systems are drawn.
The trend in recent observations appears to agree with this interpretation --- 
only after a dedicated analysis, \cite{Peters16} and \cite{Wang17} found an sdO star with uncertain luminosity as a companion to a Be star.
Moreover, \cite{Wang18} found an additional twelve candidate Be+sdO systems. The stellar parameters of the latter remain to be determined.

At this stage the sample is far too heterogeneous and poorly characterized to derive any solid conclusions, and there may be further selection effects at play. It does however provide clear motivation for a more systematic search for these elusive SESs, especially given their astrophysical significance.


\section{Summary and Conclusions \label{sec:conclusions}}

A detailed study of nearby post-interaction binary systems is one of the important ways to provide crucial insight in the uncertain physical processes of binary interaction and to help us to better understand the population of the exotic products of binary interaction.
Of particular interest is the (apparent) paradox of the scarcity of missing stripped stars.  Despite the fact that binaries are common, only very few systems are known that reside in the first long-lived post mass transfer phase. The binary system \phiper is thought to be one of these systems.  It contains a $\sim 1.2\Msun$ hot subdwarf and a $\sim 9.6\Msun$ Be star in a 126.7\, day orbit. 

We systematically reanalyze the binary system \phiper using an extensive grid of binary evolutionary models and use this to derive constraints on the progenitor, infer the current evolutionary state, and predict its future evolution. Our main findings are listed here. 
\begin{enumerate}

\item Our best fit for the progenitor of \phiper is a post-interaction binary system where the hot subdwarf is the stripped core of a star with an initial mass of $7.2\pm 0.4\Msun$.  For the rapidly rotating Be star we derive an initial mass of $3.8 \pm 0.4\Msun$. We derive an initial orbital period of $16\pm4$\,days. We find that models including conservative mass transfer, i.e., models where at least bout 75\% of the mass lost by the donor is accreted by the secondary, are favored, consistent with earlier findings. 

 \item We derive an age of $57 \pm 9$\,Myr for the system. This is in excellent agreement with the most recently determined age of its suspected host cluster $\alpha$~Persei, 52-60\,Myr. This would imply that the 9.6$\pm$0.3\Msun Be star in \phiper is a rejuvenated blue straggler, which has a turn-off mass of about 7\Msun. This also makes \phiper a likely progenitor of a late core-collapse supernova, i.e., one that explodes after all massive single stars in the cluster have already ended their lives. 
 
\item We find that the subdwarf is almost an order of magnitude brighter than expected for a helium core burning star of this mass.  It is also $\sim$3 times more luminous than the subdwarf with a similar mass in the system FY\,CMa.  
We can only reproduce the excessive luminosity with our models if the subdwarf has already completed helium core burning and is now in a more luminous phase where it is burning helium in a shell around the core.

\item The surprisingly large luminosity of the subdwarf in \phiper seems to imply that we have caught the system in a short-lived late evolutionary stage. Based upon lifetime arguments, we expect only 2-3\% of subdwarfs of this mass to be this bright. This would imply that there are up to 50 systems with birth parameters  similar to those of \phiper that contain a less evolved subdwarf which remained undetected, due to various observational biases. This appears to be an important clue for understanding the (apparent) paradox of the scarcity of detections of stripped stars in binary systems. 

\item 
We compiled an overview of all currently reported subdwarfs and helium giants with main sequence companions in literature.
The sample only yields eleven systems. Of these at least seven appear to host stripped stars that currently reside in the late stages of the short-lived helium shell burning phase. If true, this would also indicate that we are still overlooking the presence of possibly hundred(s) less evolved stripped stars. 
\end{enumerate}

Understanding the (apparent) paradox of the scarcity of detections of stripped stars in binary systems is important for various fields in astrophysics. Stripped stars have been suggested to play an important role by contributing far UV and ionizing radiation in older stellar populations. Higher mass counterparts  of \phiper's stripped star are considered to be Ib$\slash$c supernova progenitors. As such, binary systems that eventually produce gravitational wave sources through the classic formation scenario are expected to evolve through a very similar post interaction phase as \phiper occupies.  

Our results imply that a substantial population exists of systems analogous to \phiper that contain less evolved stripped stars and that have evaded detection so far.  Dedicated searches for these systems will be extremely valuable, since they can provide strong constraints on the physics of binary interaction, as we have shown.  

\begin{acknowledgements}
The authors would like to thank Rob Izzard for providing the \binaryc code, Ulrich Heber, Pablo Marchant and Philipp Podsiadlowski for help and input and Norbert Langer for useful comments on the draft. 
SdM acknowledges support by a Marie Sklodowska-Curie Action Incoming Fellowship (H2020 MSCA-IF-2014, project BinCosmos, id 661502). The authors further acknowledge the Leiden Lorentz Center for support of the workshop ``The Impact of Massive Binaries Throughout the Universe''.
\end{acknowledgements}


\bibliographystyle{aa.bst}

\begin{thebibliography}{0}
\expandafter\ifx\csname natexlab\endcsname\relax\def\natexlab#1{#1}\fi

\end{thebibliography}


\begin{thebibliography}{}

\bibitem[Almeida et al.(2017)]{Almeida17} Almeida, L.~A., Sana, H., Taylor, W., et al.\ 2017, \aap, 598, A84


\bibitem[B{\"o}hm-Vitense(1958)]{Bohm58} B{\"o}hm-Vitense, E.\ 1958, \zap, 46, 108

\bibitem[Bozic et al.(1995)]{Bozic95} Bozic, H., Harmanec, P., Horn, J., et al.\ 1995, \aap, 304, 235 

\bibitem[{{Campbell}(1902)}]{Campbell02}
{Campbell}, W.~W. 1902, \apj, 16, 114

\bibitem[Cannon(1910)]{Cannon10} Cannon, J.~B.\ 1910, \jrasc, 4, 195


\bibitem[Cantiello et al.(2007)]{Cantiello07} Cantiello, M., Yoon, S.-C., Langer, N., \& Livio, M.\ 2007, \aap, 465, L29

\bibitem[Chini et al.(2012)]{Chini12} Chini, R., Hoffmeister, V.~H., Nasseri, A., Stahl, O., \& Zinnecker, H.\ 2012, \mnras, 424, 1925



\bibitem[De Greve \& De Loore(1977)]{deGreve77} De Greve, J.-P., \& De Loore, C.\ 1977, \apss, 50, 75 

\bibitem[de Jager et al.(1988)]{deJager88} de Jager, C., Nieuwenhuijzen, H., \& van der Hucht, K.~A.\ 1988, \aaps, 72, 259

\bibitem[Delgado \& Thomas(1981)]{Delgado81} Delgado, A.~J., \& Thomas, H.-C.\ 1981, \aap, 96, 142

\bibitem[de Mink et al.(2007)]{deMink07} de Mink, S.~E., Pols, O.~R., \& Hilditch, R.~W.\ 2007, \aap, 467, 1181

\bibitem[de Mink et al.(2009)]{deMink09} de Mink, S.~E., Cantiello, M., Langer, N., et al.\ 2009, \aap, 497, 243 

\bibitem[{{de Mink} {et~al.}(2013){de Mink}, {Langer}, {Izzard}, {Sana}, \& {de Koter}}]{deMink13}
{de Mink}, S.~E., {Langer}, N., {Izzard}, R.~G., {Sana}, H., \& {de Koter}, A.
  2013, \apj, 764, 166

\bibitem[{{de Mink} {et~al.}(2014){deMink14}, {Sana}, {Langer}, {Izzard}, \&
  {Schneider}}]{deMink14}
{de Mink}, S.~E., {Sana}, H., {Langer}, N., {Izzard}, R.~G., \& {Schneider},
  F.~R.~N. 2014, \apj, 782, 7
  


\bibitem[Drilling \& Schoenberner(1982)]{Drilling82} Drilling, J.~S., \& Schoenberner, D.\ 1982, \aap, 113, L22 


\bibitem[Duch{\^e}ne \& Kraus(2013)]{Duchene13} Duch{\^e}ne, G., \& Kraus, A.\ 2013, \araa, 51, 269

\bibitem[Dudley \& Jeffery(1990)]{Dudley90} Dudley, R.~E., \& Jeffery, C.~S.\ 1990, \mnras, 247, 400 

\bibitem[Dudley \& Jeffery(1993)]{Dudley93} Dudley, R.~E., \& Jeffery, C.~S.\ 1993, \mnras, 262, 945 







\bibitem[Frame et al.(1995)]{Frame95} Frame, D.~J., Cottrell, P.~L., Gilmore, A.~C., Kilmartin, P.~M., \& Lawson, W.~A.\ 1995, \mnras, 276, 383 

\bibitem[Gies et al.(1993)]{Gies93} Gies, D.~R., Willis, 
C.~Y., Penny, L.~R., \& McDavid, D.\ 1993, \pasp, 105, 281 

\bibitem[{{Gies} {et~al.}(1998){Gies98}, {Bagnuolo}, {Ferrara}, {Kaye},
  {Thaller}, {Penny}, \& {Peters}}]{Gies98}
{Gies}, D.~R., {Bagnuolo}, Jr., W.~G., {Ferrara}, E.~C., {et~al.} 1998, ApJ, 493, 440
  

\bibitem[Graczyk et al.(2011)]{Graczyk11} Graczyk, D., Soszy{\'n}ski, I., Poleski, R., et al.\ 2011, \actaa, 61, 103

\bibitem[Groh et al.(2008)]{Groh08} Groh, J.~H., Oliveira, A.~S., \& Steiner, J.~E.\ 2008, \aap, 485, 245 

\bibitem[G{\"o}tberg et al.(2017)]{Gotberg17} G{\"o}tberg, Y., de Mink, S.~E., \& Groh, J.~H.\ 2017, \aap, 608, A11






\bibitem[Heber(2016)]{Heber16} Heber, U.\ 2016, \pasp, 128, 082001


\bibitem[{{Hurley} {et~al.}(2000){Hurley}, {Pols}, \& {Tout}}]{Hurley00}
{Hurley}, J.~R., {Pols}, O.~R., \& {Tout}, C.~A. 2000, \mnras, 315, 543

\bibitem[{{Hurley} {et~al.}(2002){Hurley}, {Tout}, \& {Pols}}]{Hurley02}
{Hurley}, J.~R., {Tout}, C.~A., \& {Pols}, O.~R. 2002, \mnras, 329, 897


\bibitem[{{Izzard} {et~al.}(2004){Izzard04}, {Tout}, {Karakas}, \&
  {Pols}}]{Izzard04}{Izzard}, R.~G., {Tout}, C.~A., {Karakas}, A.~I., \& {Pols}, O.~R. 2004, \mnras, 350, 407
  
\bibitem[{{Izzard} {et~al.}(2006){Izzard06}, {Dray}, {Karakas}, {Lugaro}, \&
  {Tout}}]{Izzard06}
{Izzard}, R.~G., {Dray}, L.~M., {Karakas}, A.~I., {Lugaro}, M., \& {Tout},
  C.~A. 2006, \aap, 460, 565

\bibitem[{{Izzard} {et~al.}(2009){Izzard09}, {Glebbeek}, {Stancliffe}, \&
  {Pols}}]{Izzard09}
{Izzard}, R.~G., {Glebbeek}, E., {Stancliffe}, R.~J., \& {Pols}, O.~R. 2009,
  A\&A, 508, 1359

\bibitem[Jeffery \& Aznar Cuadrado(2001)]{Jeffery01} Jeffery, C.~S., \& Aznar Cuadrado, R.\ 2001, \aap, 378, 936



\bibitem[Kiminki \& Kobulnicky(2012)]{Kiminki12} Kiminki, D.~C., \& Kobulnicky, H.~A.\ 2012, \apj, 751, 4 

\bibitem[Kippenhahn \& Weigert(1967)]{Kippenhahn67} Kippenhahn, R., \& Weigert, A.\ 1967, \zap, 65, 251

\bibitem[Kipper \& Klochkova(2008)]{Kipper08} Kipper, T., \& Klochkova, V.~G.\ 2008, Baltic Astronomy, 17, 195

\bibitem[Kipper \& Klochkova(2012)]{Kipper12} Kipper, T., \& Klochkova, V.~G.\ 2012, Baltic Astronomy, 21, 219


\bibitem[Kobulnicky \& Fryer(2007)]{Kobulnicky06} Kobulnicky, H.~A., \& Fryer, C.~L.\ 2007, \apj, 670, 747

\bibitem[Kouwenhoven et al.(2007)]{Kouwenhoven07} Kouwenhoven, M.~B.~N., Brown, A.~G.~A., Portegies Zwart, S.~F., \& Kaper, L.\ 2007, \aap, 474, 77


\bibitem[Kupfer et al.(2015)]{Kupfer15} Kupfer, T., Geier, S., Heber, U., et al.\ 2015, \aap, 576, A44

\bibitem[Langer et al.(1983)]{Langer83} Langer, N., Fricke, K.~J., \& Sugimoto, D.\ 1983, \aap, 126, 207 




\bibitem[Ludendorff(1910)]{Ludendorff10} Ludendorff, H.\ 1910, 
Astronomische Nachrichten, 186, 17

\bibitem[{{Makarov}(2006)}]{Makarov06}
{Makarov}, V.~V. 2006, \aj, 131, 2967

\bibitem[Mandel \& de Mink(2016)]{Mandel16} Mandel, I., \& de Mink, S.~E.\ 2016, \mnras,  

\bibitem[Marchant et al.(2016)]{Marchant16} Marchant, P., Langer, N., Podsiadlowski, P., Tauris, T.~M., \& Moriya, T.~J.\ 2016, \aap, 588, A50

\bibitem[Margoni et al.(1988)]{Margoni88} Margoni, R., Stagni, R., \& Mammano, A.\ 1988, \aaps, 75, 157



\bibitem[Massey et al.(2014)]{Massey14} Massey, P., Neugent, K.~F., Morrell, N., \& Hillier, D.~J.\ 2014, \apj, 788, 83

\bibitem[Massey et al.(2015)]{Massey15} Massey, P., Neugent, K.~F., \& Morrell, N.\ 2015, \apj, 807, 81

\bibitem[Massey et al.(2017)]{Massey17} Massey, P., Neugent, K.~F., \& Morrell, N.\ 2017, \apj, 837, 122

\bibitem[Mourard et al.(2015)]{Mourard15} Mourard, D., Monnier, J.~D., Meilland, A., et al.\ 2015, \aap, 577, A51

\bibitem[Neo et al.(1977)]{Neo77} Neo, S., Miyaji, S., Nomoto, K., \& Sugimoto, D.\ 1977, \pasj, 29, 249

\bibitem[Netopil \& Paunzen(2013)]{Netopil13} Netopil, M., \& Paunzen, E.\ 2013, \aap, 557, A10




F.-K., \& Yokoi, K.\ 1984, \apj, 286, 644


\bibitem[Nugis \& Lamers(2000)]{Nugis00} Nugis, T., \& Lamers, H.~J.~G.~L.~M.\ 2000, \aap, 360, 227

\bibitem[{{{\"O}pik}(1924)}]{Opik24}{{\"O}pik}, E. 1924, Publications of the Tartu Astrofizica Observatory, 25, 1




\bibitem[Parthasarathy et al.(1990)]{Parthasarathy90} Parthasarathy, M., Hack, M., \& Tektunali, G.\ 1990, \aap, 230, 136


\bibitem[Paxton et al.(2011)]{Paxton11}
{Paxton}, B., {Bildsten}, L., {Dotter}, A., {et~al.} 2011, \apjs, 192, 3

\bibitem[Paxton et al.(2013)]{Paxton13} Paxton, B., Cantiello, M., Arras, P., et al.\ 2013, \apjs, 208, 4

\bibitem[Paxton et al.(2015)]{Paxton15} Paxton, B., Marchant, 
P., Schwab, J., et al.\ 2015, \apjs, 220, 15 


\bibitem[{{Peters} {et~al.}(2008){Peters08}, {Gies}, {Grundstrom}, \&  {McSwain}}]{Peters08}{Peters}, G.~J., {Gies}, D.~R., {Grundstrom}, E.~D., \& {McSwain}, M.~V. 2008,
  \apj, 686, 1280
  
\bibitem[{{Peters} {et~al.}(2013){Peters13}, {Pewett}, {Gies}, {Touhami}, \& {Grundstrom}}]{Peters13}
{Peters}, G.~J., {Pewett}, T.~D., {Gies}, D.~R., {Touhami}, Y.~N., \& {Grundstrom}, E.~D. 2013, \apj, 765, 2

\bibitem[Peters et al.(2016)]{Peters16} Peters, G.~J., Wang, L., Gies, D.~R., \& Grundstrom, E.~D.\ 2016, \apj, 828, 47

\bibitem[Petrovic et al.(2005)]{Petrovic05} Petrovic, J., Langer, N., Yoon, S.-C., \& Heger, A.\ 2005, \aap, 435, 247

  W.~{van Rensbergen}, \& C.~{De Loore}, 439--453

\bibitem[Podsiadlowski et al.(1992)]{Podsiadlowski92} Podsiadlowski, P., Joss, P.~C., \& Hsu, J.~J.~L.\ 1992, \apj, 391, 246
  


\bibitem[Poeckert(1979)]{Poeckert79} Poeckert, R.\ 1979, \apjl, 233, L73 

\bibitem[{{Poeckert}(1981)}]{Poeckert81}{Poeckert}, R. 1981, \pasp, 93, 297


\bibitem[{{Pols} {et~al.}(1991){Pols91}, {Cote}, {Waters}, \& {Heise}}]{Pols91}
{Pols}, O.~R., {Cote}, J., {Waters}, L.~B.~F.~M., \& {Heise}, J. 1991, A\&A,
  241, 419
  
\bibitem[Pols(1994)]{Pols94} Pols, O.~R.\ 1994, \aap, 290, 119
  
\bibitem[{{Pols} {et~al.}(1998){Pols}, {Schr{\"o}der}, {Hurley}, {Tout}, \&
  {Eggleton}}]{Pols98}
{Pols}, O.~R., {Schr{\"o}der}, K.-P., {Hurley}, J.~R., {Tout}, C.~A., \&
  {Eggleton}, P.~P. 1998, \mnras, 298, 525

\bibitem[Pols(2007)]{Pols07} Pols, O.~R.\ 2007, in Massive Stars in Interactive Binaries, ASP Conf. Ser. 367, ed. N. St-Louis \& A. F. J. Moffat (San Francisco: ASP), 387



\bibitem[Ritter(1988)]{Ritter88} Ritter, H.\ 1988, \aap, 202, 93 

\bibitem[{{Salpeter}(1955)}]{Salpeter55}{Salpeter}, E.~E. 1955, \apj, 121, 161

\bibitem[{{Sana} {et~al.}(2012){Sana12}, {De Mink}, {de Koter}, {Langer},
  {Evans}, {Gieles}, {Gosset}, {Izzard}, {Le Bouquin}, \& {Schneider}}]{Sana12}
{Sana}, H., {De Mink}, S.~E., {de Koter}, A., {et~al.} 2012, Science, 337, 444


\bibitem[Schoenberner \& Drilling(1983)]{Schoenberner83} Schoenberner, D., \& Drilling, J.~S.\ 1983, \apj, 268, 225

\bibitem[Schoenberner \& Drilling(1984)]{Schoenberner84} Schoenberner, D., \& Drilling, J.~S.\ 1984, \apj, 276, 229 

\bibitem[Slettebak(1982)]{Slettebak82} Slettebak, A.\ 1982, \apjs, 50, 55 


\bibitem[Smith et al.(2018)]{Smith17} Smith, N., G{\"o}tberg, Y., \& de Mink, S.~E.\ 2018, \mnras, 475, 772

\bibitem[Soberman et al.(1997)]{Soberman97} Soberman, G.~E., Phinney, E.~S., \& van den Heuvel, E.~P.~J.\ 1997, \aap, 327, 620

\bibitem[Stancliffe \& Glebbeek(2008)]{Stancliffe08} Stancliffe, R.~J., \& Glebbeek, E.\ 2008, \mnras, 389, 1828 



\bibitem[Stauffer et al.(1999)]{Stauffer99} Stauffer, J.~R., 
Navascu{\'e}s, D.~B.~y., Bouvier, J., et al.\ 1999, \apj, 527, 219 


\bibitem[Tauris \& van den Heuvel(2006)]{Tauris06} Tauris, T.~M., \& van den Heuvel, E.~P.~J.\ 2006, Compact stellar X-ray sources, 623 

\bibitem[Thaller et al.(1995)]{Thaller95} Thaller, M.~L., Bagnuolo, W.~G., Jr., Gies, D.~R., \& Penny, L.~R.\ 1995, \apj, 448, 878 




\bibitem[Vanbeveren et al.(1998a)]{Vanbeveren98} Vanbeveren, D., De Loore, C., \& Van Rensbergen, W.\ 1998, \aapr, 9, 63 

\bibitem[Vanbeveren et al.(1998b)]{Vanbeveren98b} Vanbeveren, D., van Rensbergen, W., \& De Loore, C.\ 1998, Astrophysics and Space Science Library, 232


\bibitem[van den Heuvel et al.(2017)]{vandenHeuvel17} van den Heuvel, E.~P.~J., Portegies Zwart, S.~F., \& de Mink, S.~E.\ 2017, \mnras, 471, 4256



\bibitem[Vink et al.(2001)]{Vink01} Vink, J.~S., de Koter, A., \& Lamers, H.~J.~G.~L.~M.\ 2001, \aap, 369, 574

\bibitem[von Zeipel(1924)]{vonZeipel24} von Zeipel, H.\ 1924, \mnras, 84, 665


\bibitem[Wang et al.(2017)]{Wang17} Wang, L., Gies, D.~R., \& Peters, G.~J.\ 2017, \apj, 843, 60

\bibitem[Wang et al.(2018)]{Wang18} Wang, L., Gies, D.~R., \& Peters, G.~J.\ 2018, \apj, 853, 160


\bibitem[Waters(1986)]{waters86} Waters, L.~B.~F.~M.\ 1986, \aap, 162, 121


\bibitem[Wellstein et al.(2001)]{Wellstein01} Wellstein, S., Langer, N., \& Braun, H.\ 2001, \aap, 369, 939

\bibitem[Yoon et al.(2010)]{Yoon10} Yoon, S.-C., Woosley, S.~E., \& Langer, N.\ 2010, \apj, 725, 940 


\bibitem[Yoon et al.(2017)]{Yoon17} Yoon, S.-C., Dessart, L., \& Clocchiatti, A.\ 2017, arXiv:1701.02089

\bibitem[Zapartas et al.(2017)]{zapartas17} Zapartas, E., de Mink, S.~E., Izzard, R.~G., et al.\ 2017, arXiv:1701.07032

\bibitem[{{Zuckerman} {et~al.}(2012){Zuckerman}, {Melis}, {Rhee}, {Schneider}, \& {Song}}]{Zuckerman12}{Zuckerman}, B., {Melis}, C., {Rhee}, J.~H., {Schneider}, A., \& {Song}, I. 2012, \apj, 752, 58

\end{thebibliography}

\end{document}